\documentclass[12pt,english]{article}
\pdfoutput=1
\usepackage[T1]{fontenc}
\usepackage{graphicx,array}
\usepackage{cite}
\usepackage{color}
\usepackage{latexsym}
\usepackage{amsthm}
\usepackage{amsmath}
\usepackage[titletoc]{appendix}
\usepackage{enumitem}
\usepackage{amssymb}
\usepackage{color}
\usepackage{float}
\usepackage{outlines}
\usepackage[unicode=true,
 bookmarks=true,bookmarksnumbered=false,bookmarksopen=false,
 breaklinks=false,pdfborder={0 0 1},backref=false,colorlinks=true]
 {hyperref}
\hypersetup{pdftitle={Holevo Information of Black Hole Mesostates},
 pdfauthor={Ning Bao, Jonathan Harper, Grant N. Remmen},
 citecolor=black,linkcolor=black,urlcolor=black}
\usepackage{breakurl}
\usepackage[hang,flushmargin]{footmisc} 

\def\simgt{\mathrel{\lower2.5pt\vbox{\lineskip=0pt\baselineskip=0pt
           \hbox{$>$}\hbox{$\sim$}}}}
\def\simlt{\mathrel{\lower2.5pt\vbox{\lineskip=0pt\baselineskip=0pt
           \hbox{$<$}\hbox{$\sim$}}}}

\newcommand{\be}{\begin{equation}}
\newcommand{\ee}{\end{equation}}
\newcommand{\bea}{\begin{eqnarray}}
\newcommand{\eea}{\end{eqnarray}}
\newcommand{\Eq}[1]{Eq.~(\ref{#1})}

\newcommand{\Sec}[1]{Sec.~\ref{#1}}

\newcommand{\Fig}[1]{Fig.~\ref{#1}}

\newcommand*\oline[1]{%
  \vbox{%
    \hrule height 0.5pt
    \kern0.68ex
    \hbox{%
      \kern-0.1em
      \ifmmode#1\else\ensuremath{#1}\fi
      \kern-0.1em
    }
  }
}

\definecolor{nicered}{rgb}{0.7,0.1,0.1}
\definecolor{nicegreen}{rgb}{0.1,0.5,0.1}
\hypersetup{
 colorlinks,citecolor=black,linkcolor=black,urlcolor=black}
\usepackage{xcolor}

\begin{document}

\interfootnotelinepenalty=10000
\baselineskip=18pt
\hfill
BRX-TH-6674

\vspace{1.5cm}
\thispagestyle{empty}
\begin{center}
{\LARGE \bf
Holevo Information of Black Hole Mesostates
}\\
\bigskip\vspace{5mm}{
{\large Ning Bao,${}^{a}$ Jonathan Harper,${}^{b}$ and Grant N. Remmen${}^{c,d}$}
} \\[5mm]
{\it ${}^a$Computational Science Initiative, Brookhaven National Lab, Upton, NY, 11973 \\[1.5mm]
${}^b$Martin Fisher School of Physics, Brandeis University, Waltham, MA 02453 \\[1.5mm]
${}^c$Kavli Institute for Theoretical Physics, University of California, Santa Barbara, CA 93106 \\
${}^d$Department of Physics, University of California, Santa Barbara, CA 93106
}
\let\thefootnote\relax\footnote{e-mail: 
\url{ningbao75@gmail.com}, \url{jharper@brandeis.edu}, \url{remmen@kitp.ucsb.edu}}
\end{center}

\bigskip
\centerline{\large\bf Abstract}
\begin{quote} \small
We define a bulk wormhole geometry interpolating between horizons of differing size and determine characteristics of the HRT surface in these geometries. This construction is dual to black hole mesostates, an intermediate coarse-graining of states between black hole microstates and the full black hole state. We analyze the distinguishability of these objects using the recently-derived holographic Holevo information techniques, demonstrating novel phase transition behavior for such systems.
\end{quote}
	
\setcounter{footnote}{0}

\newpage
\tableofcontents
\newpage

\section{Introduction}\label{sec:intro}
Distinguishing black hole microstates from each other is a necessary prerequisite to a microscopic counting of these states~\cite{Strominger_1996}. Understanding how to microscopically count black hole microstates is essential to interpolating between black hole thermodynamics and black hole statistical mechanics. Recently, in Ref.~\cite{Bao_2017}, progress was made in understanding this distinguishability task based on a quantum information theoretic quantity known as the Holevo information~\cite{holevo1973bounds}. Further work along this direction was performed in Ref.~\cite{Michel_2018}.

Despite this progress, an open question is whether one needs to go all the way to the ultraviolet to study black hole physics beyond black hole thermodynamics. In this work, we study a generalization of this question from the accounting of black hole microstates to that of {\it black hole mesostates}, as originally defined in Ref.~\cite{sorkin1999large}. Black hole mesostates are distinct ensembles of microstates of fixed ensemble entropy. That is, we are ``block renormalizing'' the microstates together, i.e., for some region $A$ of the boundary described by reduced density matrix $\rho_A$, we take subsums of the form $\bar\rho_{A,(n,m)}=\sum_{i=n}^{n+m} p_i \rho_{A,i}$, which we index as $\bar \rho_{A,j}$, where $\rho_{A,i}$ are the purified density matrices describing the microstates. These sums can be constructed such that the $\bar\rho_{A,j}$ all have the same entanglement entropy and, by a simple reorganization of the original sum, such that $\rho_A=\sum_j \bar p_j \bar\rho_{A,j}$. So long as these entropies are small enough, more fine-grained information about the statistical mechanics of the black hole can be understood with a more complete understanding of such mesostates, as they function as an interpolation between microstates and the full black hole mixed state. 

For the purposes of this paper, we will choose a grouping of the microstates into mesostates such that each of the mesostates $\bar \rho_{A,j}$ is, on its own, dual to a bulk that is geometric. Thus, here our mesostates will be thought of as geometries that act as interpolations between the thermofield double state, which corresponds to the trivial mesostate partitioning, and the black hole microstates. This interpolation can be thought of as a coarse-graining of the black hole microstates.

The question of how to treat such a coarse-graining in the context of quantum gravity, particularly in entropic calculations in holography, is an active area of current research.
This open question motivates our investigation into the intermediate microstate coarse-grainings given by the black hole mesostates.
In this paper, the geometric region of the bulk spacetime described by our chosen mesostates is strictly larger than, and indeed contains, the entanglement wedge of the original black hole, but nonetheless possesses a horizon (i.e., the mesostate is not itself pure).
While one could choose alternative fine-grainings of the original entanglement wedge, e.g., groupings of the microstates that do not possess such an interpretation, we choose to focus on these special, geometric mesostates, so that we can apply the technology of classical general relativity in order to analyze their behavior.
In effect, we will be studying partial purifications of the mixed states described by black holes; such partial purifications will simply happen to be described by classical geometry in a particular region, and we study their behavior subject to various geometric constraints.
Indeed, classical gravity has been shown to be of relevance in various contexts for understanding holographic renormalization group flow~\cite{Engelhardt_2018,Nomura_2018,Bousso_2019,Bousso:2015qqa}, which is one of the main motivations for our present investigation of black hole mesostates.

This work will be organized as follows. In Sec.~\ref{sec:Holevo}, we will review the relevant information theoretic  background related to the Holevo information. In Sec.~\ref{sec:micro}, we will discuss the work of Ref.~\cite{Bao_2017} as it relates to black hole microstates, specifically its framing of black hole microstate distinguishability through the lens of the Holevo information in asymptotically-AdS spacetimes. 
We next introduce the concept of {\it subsystem outer entropy} in Sec.~\ref{sec:subsystem}, generalizing the work on outer entropy in Refs.~\cite{Nomura_2018,Bousso_2019,Engelhardt_2018} to black hole mesostates: while the outer entropy is the largest Hubeny-Rangamani-Takayanagi~(HRT) \cite{Hubeny:2007xt} surface compatible with a fixed causal wedge exterior to some surface, the subsystem outer entropy computes the outer entropy with the additional restriction that the spacetime {\it inside} the original surface must contain some fixed geometric features (e.g., another marginally-trapped surface). 
In the case of two marginally-trapped surfaces, we prove a formula that explicitly computes the subsystem outer entropy.
In Sec.~\ref{sec:meso}, we will generalize the work of Ref.~\cite{Bao_2017} to black hole mesostates. Next, in Sec.~\ref{sec:multiboundary} we will describe a wormhole geometry motivated by Refs.~\cite{Engelhardt_2018,Bousso_2019,Nomura_2018} that reflects some of the features of black hole mesostates. Finally, in Sec.~\ref{sec:discussion} we will conclude with discussion of potential future directions, including a wormhole/tensor network-based approach to potentially access statistical mechanical features of black hole micro/mesostates.

\section{Holevo Information}\label{sec:Holevo}
Let us first briefly review the key properties of the Holevo information \cite{holevo1973bounds}. The general definition involves a distinguishability task between two players, Alice and Bob. Both players know a specific density matrix $\rho$ and moreover know the $p_i$ and $\rho_i$ that combine to make $\rho=\sum_i p_i \rho_i$. The $\rho_i$ are permitted to be pure or mixed. Without telling Bob which, Alice selects a specific $\rho_{X}$ among the $\rho_i$ and passes it to Bob. It is then Bob's task to determine the value of $X$, i.e., distinguish $\rho_{X}$ from the other $\rho_i$ via the application of a specific measurement $Y$. His ability to do this, optimized over all measurement choices, is known as the \textit{accessible information} $\sup_{Y}I(X:Y)$, where $I$ is the mutual information between the identity of the system $X$ chosen and the measurement outcome $Y$. This accessible information is upper bounded by the Holevo information $\chi$, which is given by 
\begin{equation}
    \chi(\rho,\rho_i, p_i)\equiv S(\rho)-\sum_i p_i S(\rho_i).
\end{equation}
The accessible information equals the Holevo information when all of the $\rho_i$ commute with each other. This is a situation known as \textit{optimal distinguishability}. Optimal distinguishability occurs if and only if equality between the accessible information and the Holevo information holds. Furthermore, both quantities become equal to the Shannon entropy $-\sum_i p_i \log p_i$ if the $\rho_i$ are simultaneously diagonalizable. As one cannot have better distinguishability in the abstract than the Shannon entropy  of a mixture of states, this is a situation appropriately called \textit{perfect distinguishability}. In this case, perfect distinguishability is achieved via projective measurement in the simultaneously diagonalizing basis. Similarly to optimal distinguishability, perfect distinguishability holds if and only if the Holevo information equals the Shannon entropy.

\section{Black Hole Microstates}\label{sec:micro}
Let us now specialize to the case of black hole microstates, in the context of the gauge/gravity correspondence. For a more comprehensive review, see Ref.~\cite{Bao_2017}. Throughout, we will consider bulk spacetimes that are smooth and satisfy the null energy condition. Consider some region $A$ in the boundary CFT. As before, we write $\rho_A$ for the reduced density matrix corresponding to $A$ in the boundary CFT state dual to an AdS-Schwarzschild black hole mixed state, where the density matrix of the entire boundary CFT has entanglement entropy equal to the area of the event horizon, and define $\rho_{A,i}$ to be the reduced density matrices of $A$ associated with geometries that are identical to AdS-Schwarzschild outside of the original horizon region, but which are (partially) purified by the black hole microstates inside the original location of the horizon. Importantly, we are permitted to choose the state of the entire boundary associated with these microstate geometries to be pure, as required by the definition of microstates.

From this setup, we can compute the Holevo information corresponding to this ensemble of states, leveraging the Ryu-Takayanagi (RT) formula. For simplicity, let us consider the situation where $A$ is an entire connected component of the boundary, in which case $S(\rho_A)$ is simply given by $S_{\rm BH}$, while the $S(\rho_{A,i})$ are all zero due to the purity of the microstates. (Throughout, we take $S_{\rm BH}$ to be given by the usual Bekenstein-Hawking formula, i.e., the black hole macrostate dual to $A$ is not described by some small collection of black hole microstates.) In this case, therefore, the Holevo information is equal to the Shannon entropy, and one has perfect ability to distinguish between the microstates.

One can also analyze the Holevo information for other sizes of $A$. Here we will specialize to the case where $A$ is a single interval in ${\rm AdS}_3/{\rm CFT}_2$, though the results will generalize in straightforwardly to higher dimensions.

\begin{figure}[t]
\centering
\includegraphics[width=.5\textwidth]{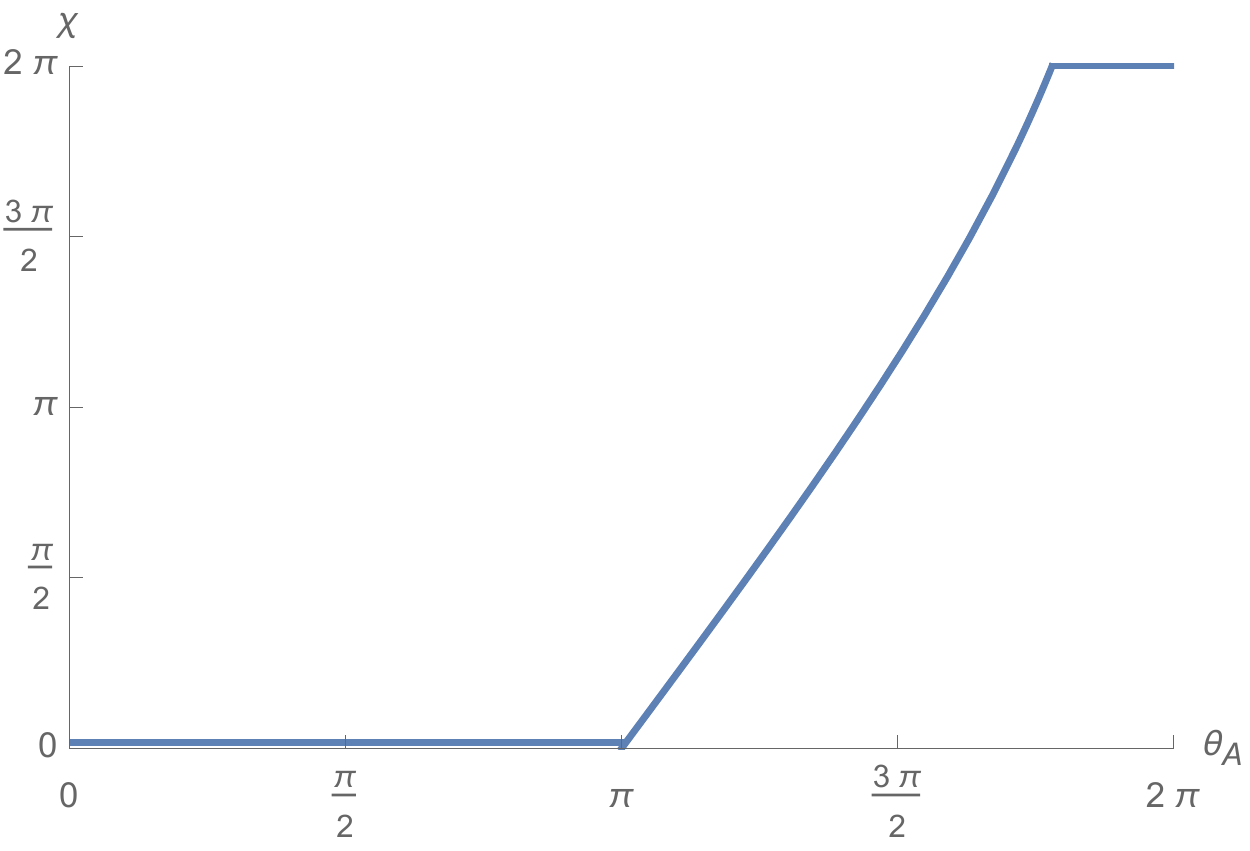} 
\caption{\label{fig:chi_BH} The Holevo information between a black hole and its microstates, $\chi_{\rm BH|micro}$, as a function of region size $\theta_A$.}
\end{figure}

\begin{figure}[t]
\centering
\includegraphics[width=.25\textwidth]{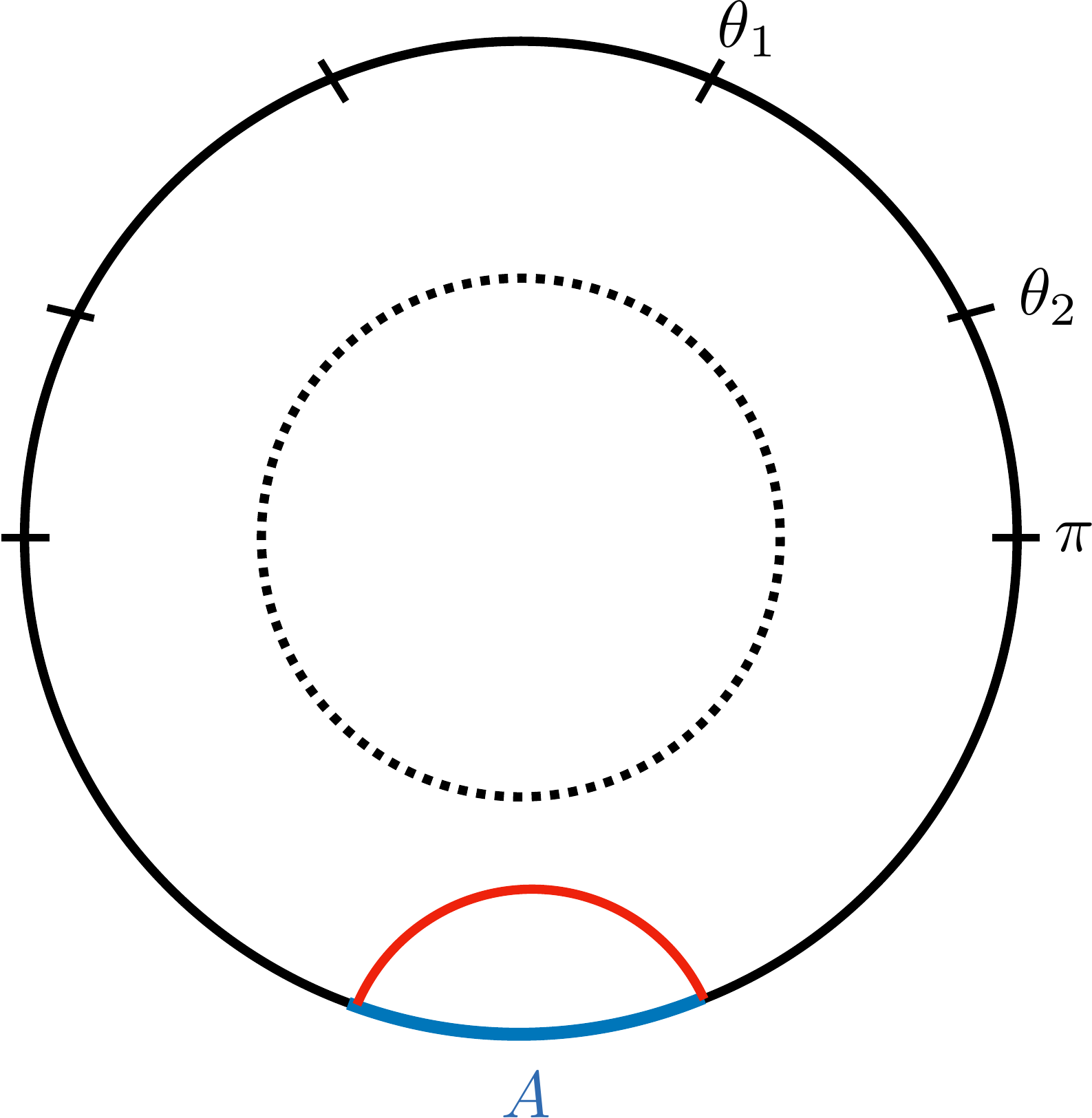}\quad \includegraphics[width=.25\textwidth]{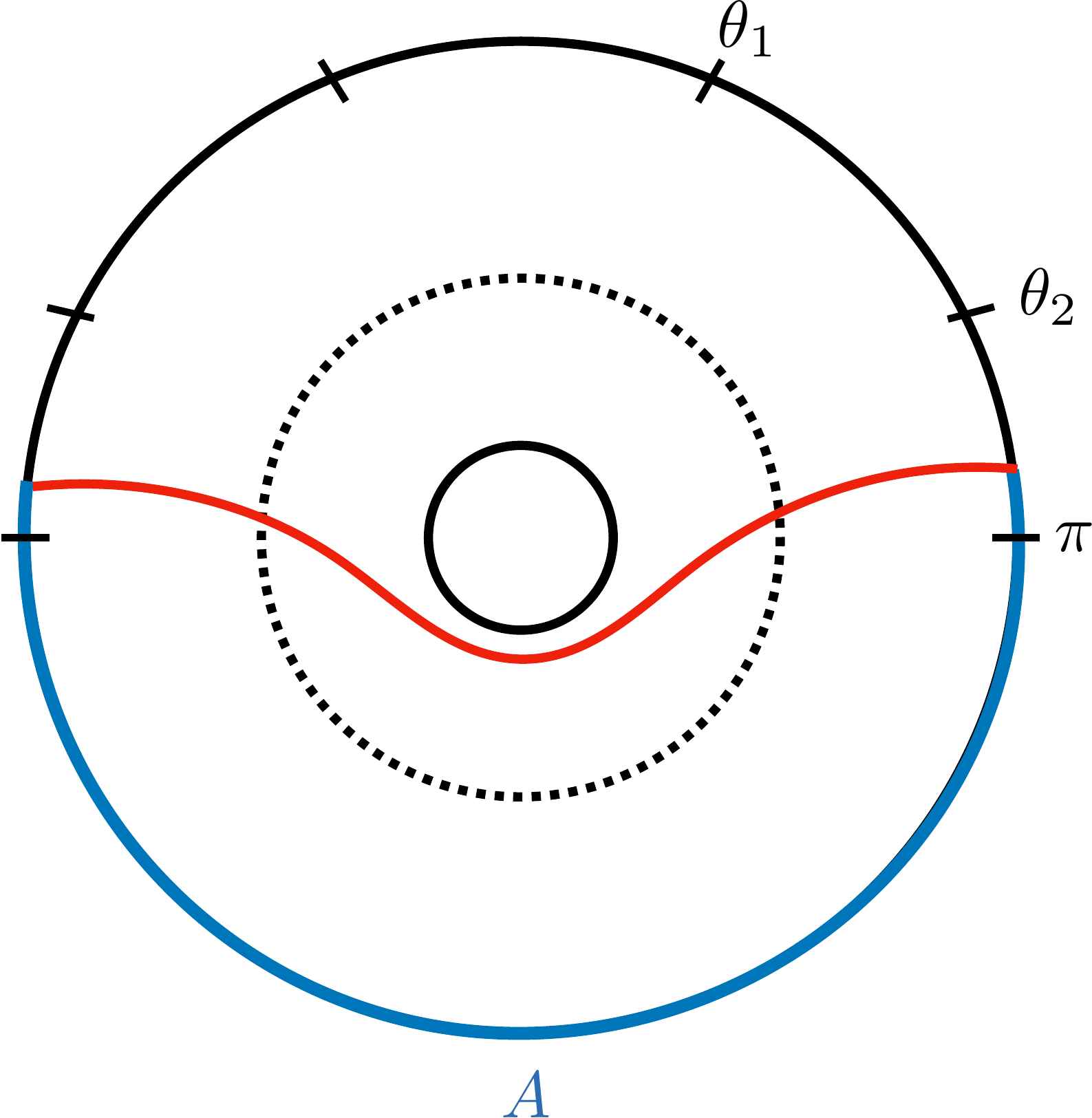}\quad \includegraphics[width=.25\textwidth]{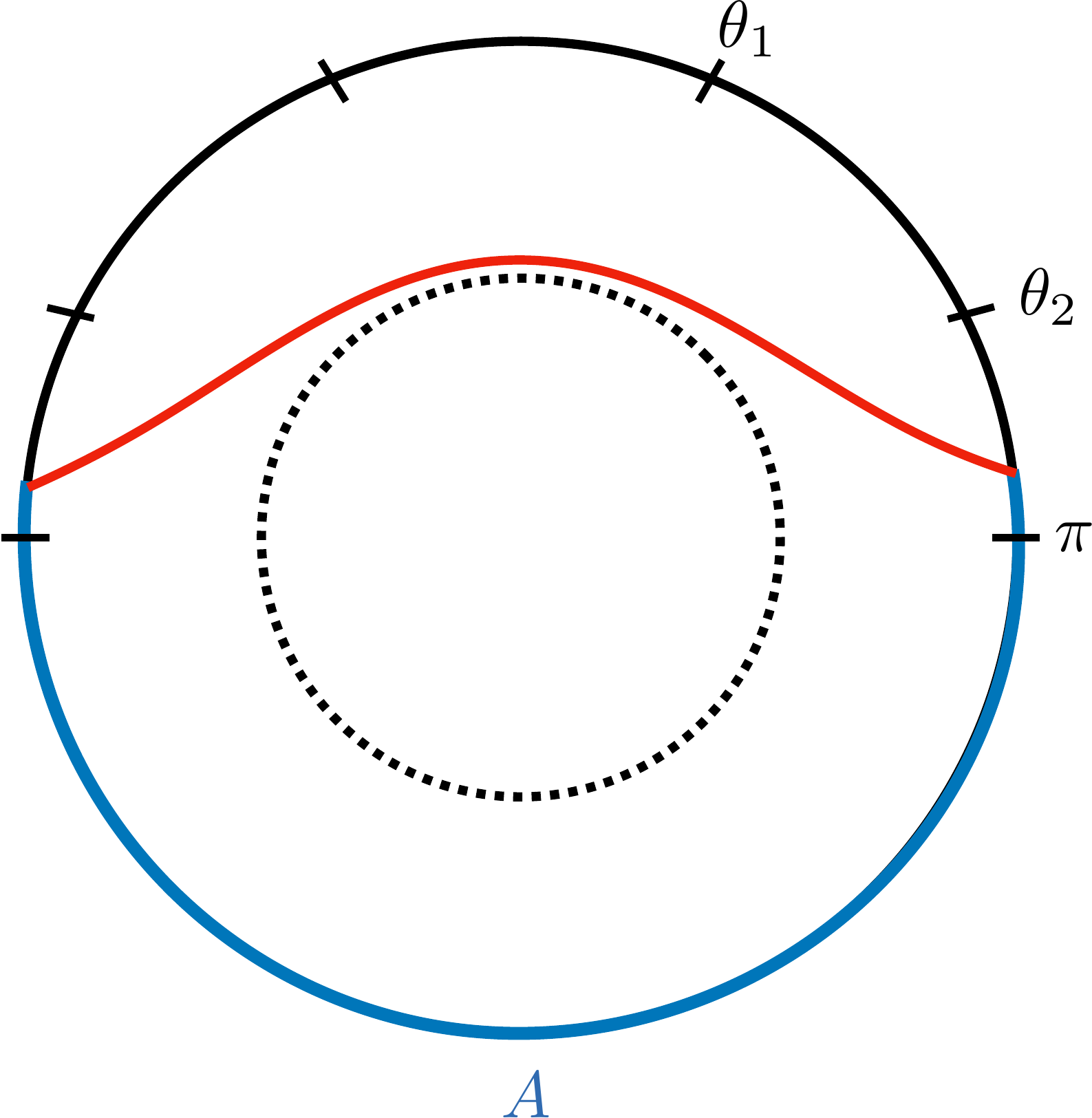}
\caption{\label{fig:micro} The microstate geometry has a phase transition at $\theta_A=\pi$ as the minimal surface switches sides. There is no thermal contribution. The coordinates $\theta_{1,2}$, associated with other phase transitions in black hole spacetimes, are defined in the captions of Figs.~\ref{fig:BH} and \ref{fig:meso} and are included throughout for purposes of comparison.}
\end{figure}

\begin{figure}[t]
\centering
\includegraphics[width=.25\textwidth]{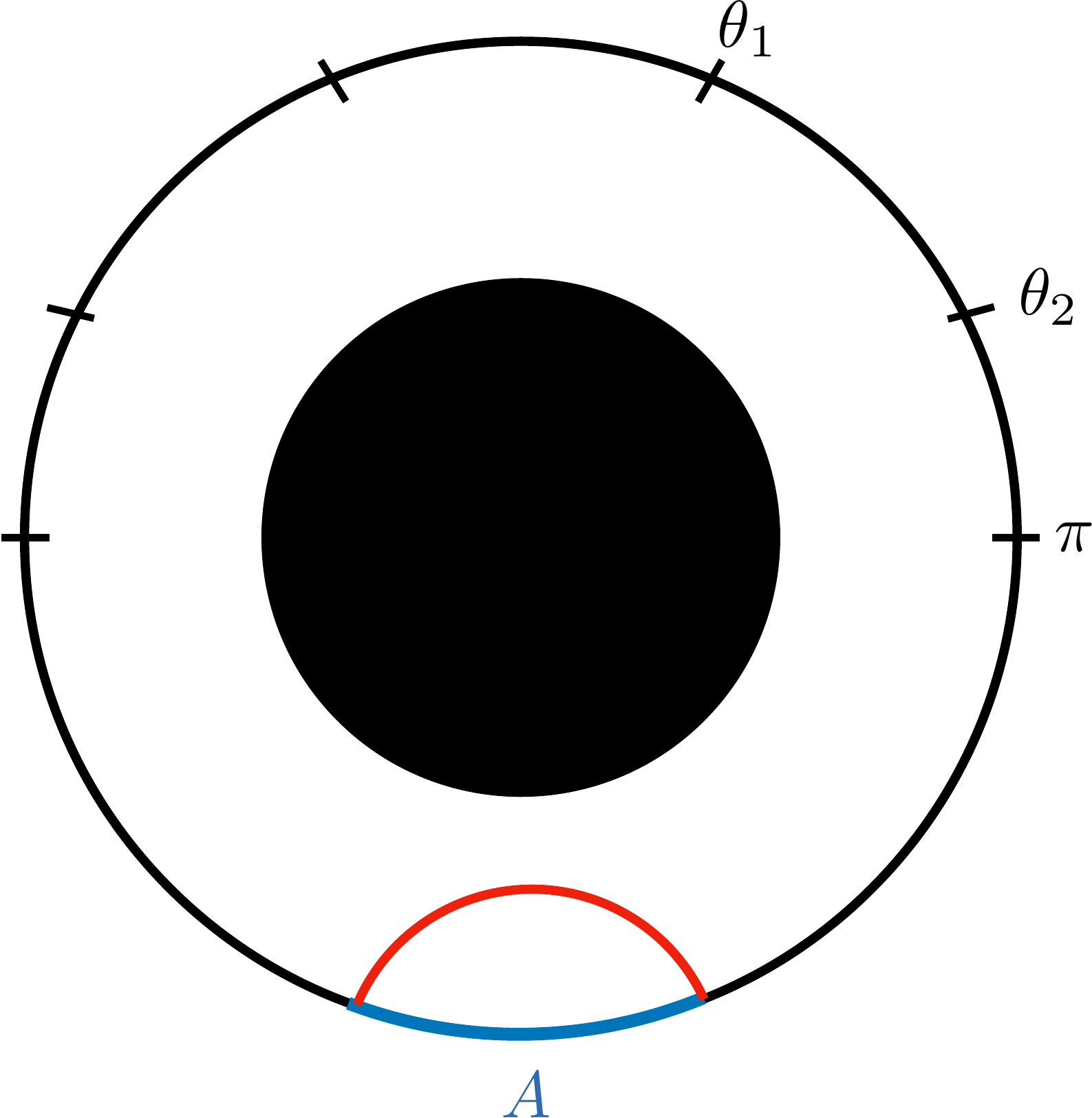} \quad \includegraphics[width=.25\textwidth]{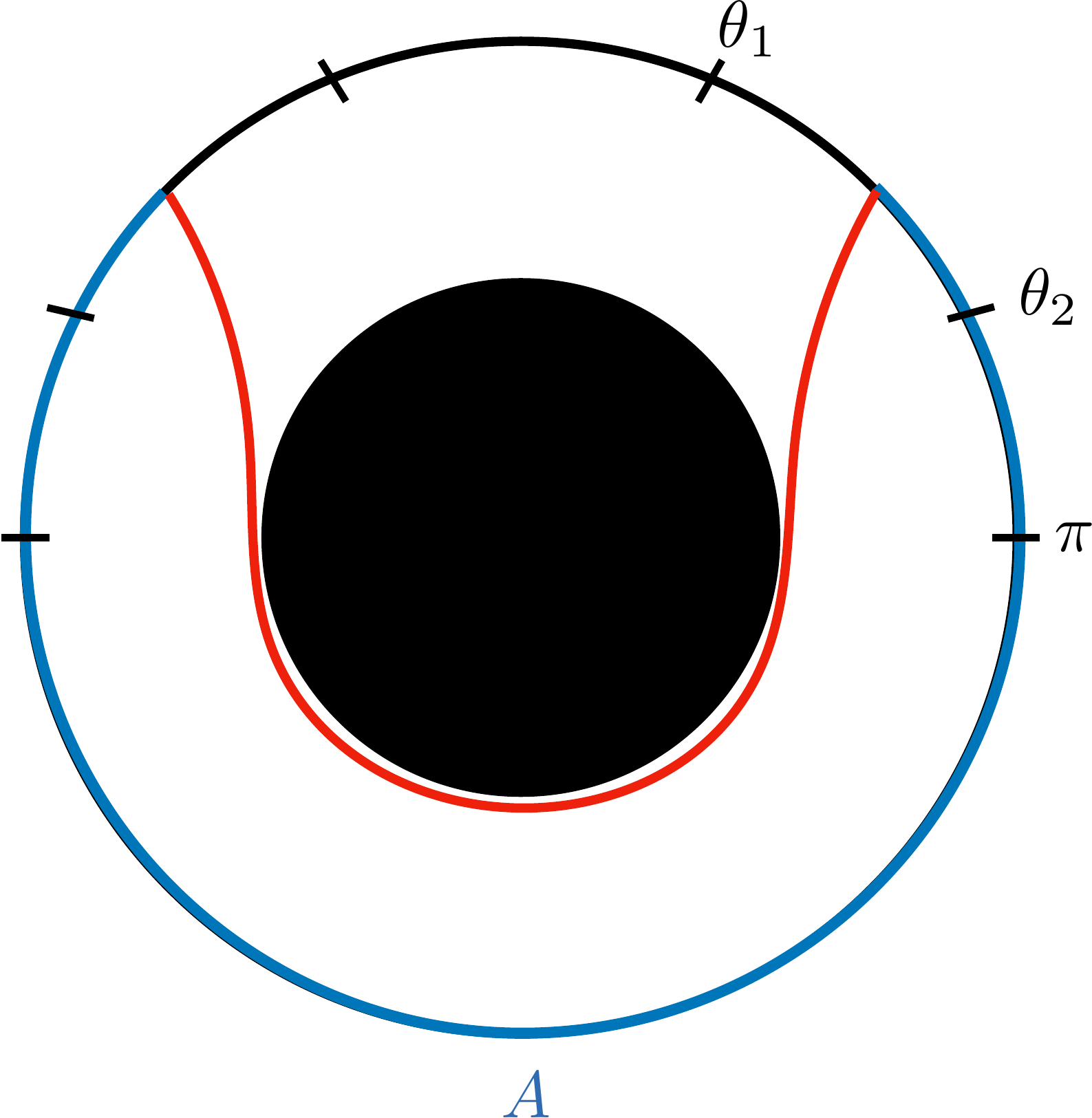} \quad \includegraphics[width=.25\textwidth]{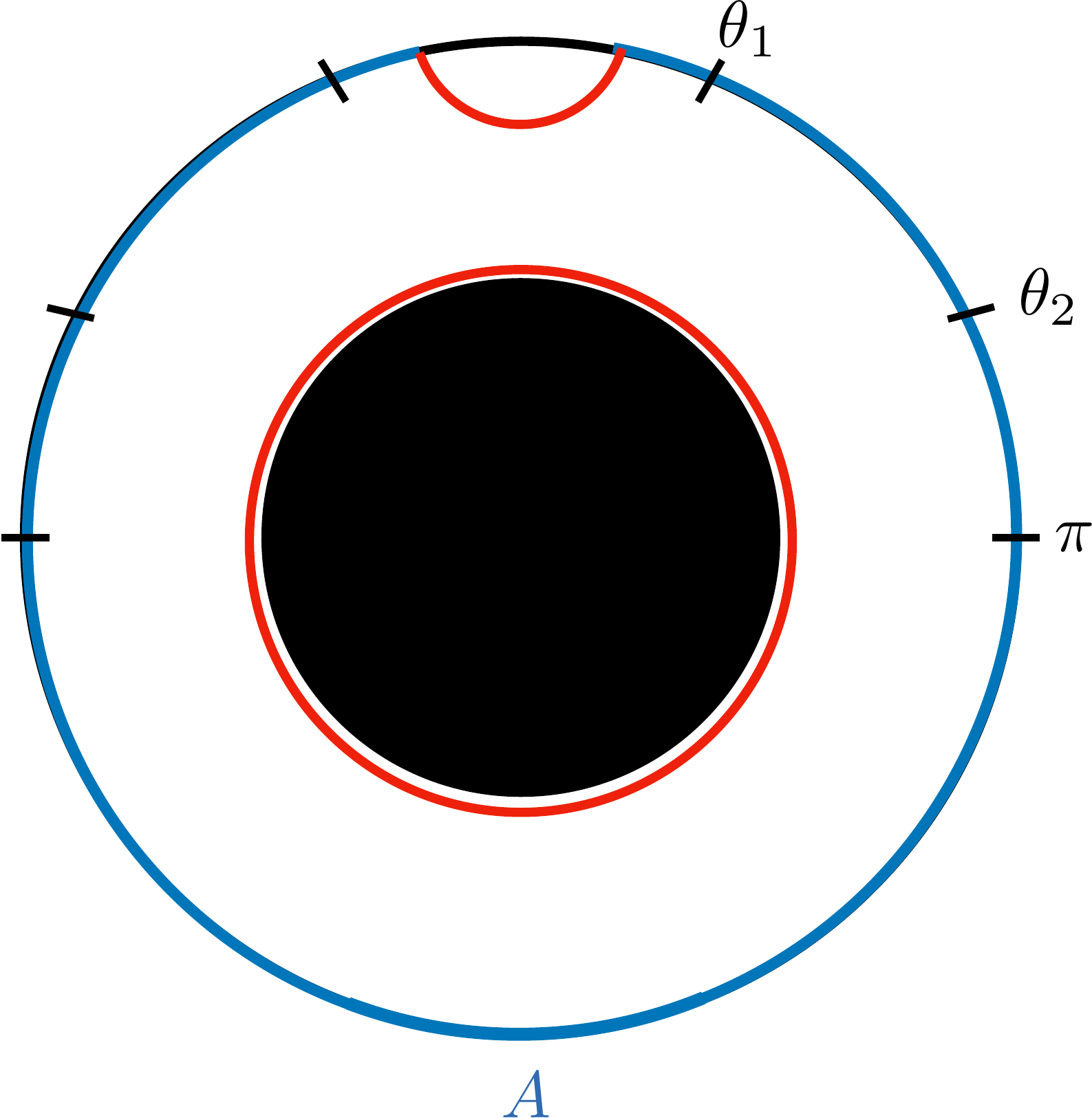}
\caption{\label{fig:BH} The black hole geometry undergoes a phase transition when the coordinate size of $A$, $\theta_A$, equals $\theta_1$. The homology condition requires the contribution from the black hole horizon at radius $r_1$, depicted by the black disk.}
\end{figure}

The subsystem entanglement $S(\rho_{A,i})$ is always given by $\min[S(\rho_A),S(\rho_{\bar{A}} )]$, as the microstates are pure state geometries and thus the horizon does not provide a homology constraint. On the other hand, $S(\rho_A)$ must respect the homology constraint imposed by the black hole horizon. This constraint is trivial when $A$ is smaller than half of the boundary CFT, and so in this range $\chi=0$. Above this value, $\chi=S(\rho_A)-\sum_i p_i S(\rho_{A,i})$ monotonically increases until the RT phase transition where the RT surface becomes disconnected, at which point it saturates at $S_{\rm BH}$; see Figs.~\ref{fig:chi_BH},~\ref{fig:micro}, and \ref{fig:BH} for an illustration. In these systems, therefore, perfect distinguishability---in the sense of a projective measurement in the diagonalized basis, as discussed in \Sec{sec:Holevo}---over the black hole microstates is attained only when one has sufficient access to the boundary so as to be past the RT phase transition.


\section{Subsystem Outer Entropy}\label{sec:subsystem}

We wish to generalize the consideration of holographic distinguishability to partial purifications, i.e., to mesostates.
Quantum mechanically, one can think of a partial purification of some reduced density matrix as the specification of a subset of the degrees of freedom purifying the state.
To remain within a geometric framework, the mixed state of the boundary CFT corresponds to the region in the causal wedge outside of a horizon, while a purification in this context means the extension of this outer wedge to a complete geometry behind the horizon and to other, disconnected, asymptotically-AdS regions describing CFTs with which our state is entangled (analogous to the thermofield double purification of the one-sided black hole-AdS geometry to a two-sided AdS wormhole).
A partial purification in this context means the specification of {\it some} geometric data behind the horizon.
Given some fixed geometric data, we can then ask questions such as: What is maximum entanglement that $A$ can have with its purification (viewing $A$ and its purification as constituent subsystems of a larger system), subject to this behind-the-horizon constraint? 

Concretely, suppose we have two apparent horizons $\sigma_1$ and $\sigma_2$ in two asymptotically-AdS geometries $M_1$ and $M_2$, respectively.\footnote{Note that we have not assumed that either of the $\sigma_i$ is itself simply connected; for example, if $M_1$ consists of multiple disconnected asymptotically-AdS spacetimes, then $\sigma_1$ would also be the disjoint union of multiple apparent horizons. Moreover, $M_1$ and $M_2$ are a priori {\it not} regions of the same connected geometry; our task will eventually be to construct a spacetime that connects their respective outer wedges.}
To be precise, by {\it apparent horizon}, we mean that we take $\sigma_1$ and $\sigma_2$ to be codimension-two surfaces that are {\it outermost} and {\it marginally trapped}.
Defining the two future-directed orthogonal null congruences on $\sigma_i$, $k_\pm[\sigma_i]$, where $k_+$ points toward $\partial M_i$, we have the null expansions $\theta_\pm[\sigma_i] = \nabla_\mu k_\pm^\mu[\sigma_i]$; the requirement of marginal trappedness is simply that $\theta_+[\sigma_i]=0$, and we further impose the generic conditions $\theta_-[\sigma_i]<0$ and $k_{+}^\mu[\sigma_i]\nabla_\mu \theta_- [\sigma_i] < 0$.
By outermost, we mean that $\sigma_i$ is homologous with $\partial M_i$ and there exists a (not necessarily unique) partial Cauchy surface $\Sigma_i$ connecting $\sigma_i$ with the boundary such that any surface within $\Sigma_i$ circumscribing $\sigma_i$ has area strictly greater than $\sigma_i$.

\begin{figure}[t]
\centering
\includegraphics[width=.45\textwidth,page=1]{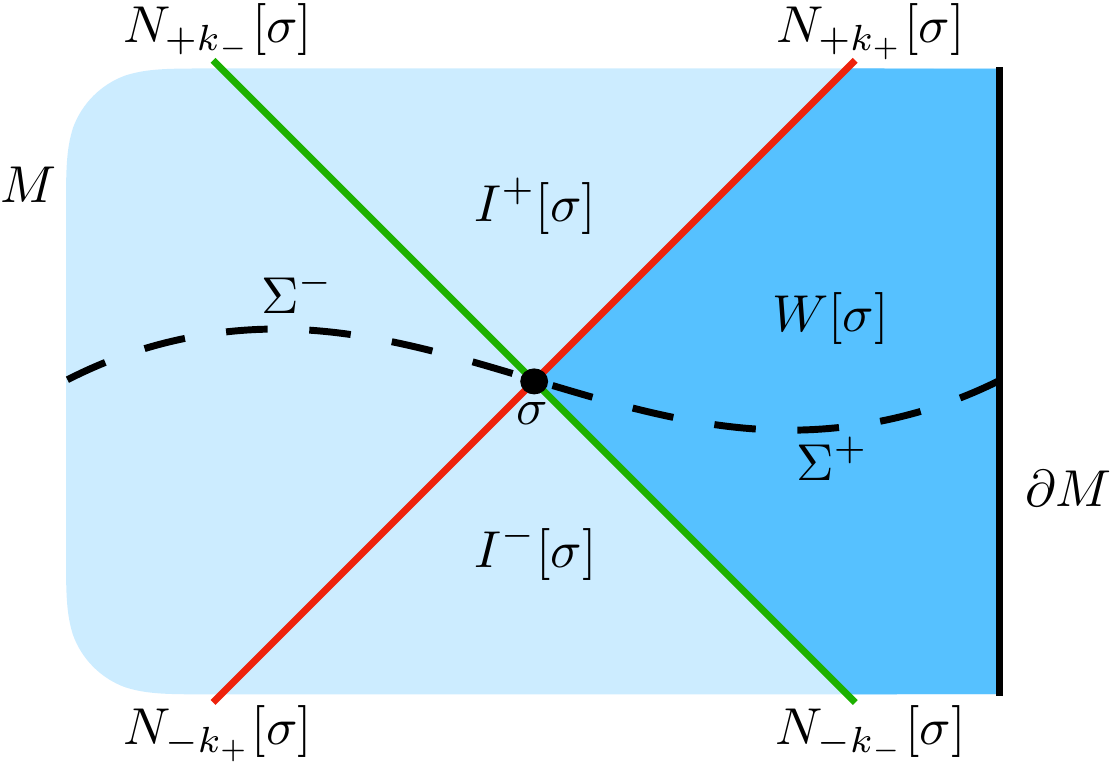}
\caption{Light sheets for a Cauchy-splitting surface $\sigma$ as defined in \Eq{eq:lightsheets}.\label{fig:lightsheets}}  
\end{figure}

Before proceeding, it will be useful to state some definitions; see Refs.~\cite{Nomura_2018,Bousso_2019, Engelhardt_2018,Bousso:2015qqa}. Given a set of spacetime points $S$, let us define the future (respectively, past) domains of dependence $D^\pm[S]$ as the set of all points from which all past (respectively, future) inextendible causal curves must contain some point in $S$, and write $D[S]\equiv D^+[S]\cup D^-[S]$.
In an asymptotically-AdS spacetime $M$, given a Cauchy surface $\Sigma$ split into two pieces $\Sigma_\pm$ by a codimension-two surface $\sigma$, where $\Sigma_+$ is connected to $\partial M$, let us define the light sheets emanating from $\sigma$: 
\begin{equation} 
\begin{aligned}
N_{\pm k_+}[\sigma] &\equiv \dot I^\pm [\Sigma^\mp] - \Sigma^\mp \\
N_{\pm k_-}[\sigma] &\equiv \dot I^\pm [\Sigma^\pm] - \Sigma^\pm,
\end{aligned}\label{eq:lightsheets}
\end{equation}
where $I^\pm[S]$ denotes the chronological future or past of a set $S$ and a dot denotes its boundary.
A light sheet is swept out by the parallel transport of the appropriate orthogonal null congruence anchored to $\sigma$, terminating at caustics or nonlocal self-intersections~\cite{Akers:2017nrr}.
We further define the outer wedge $W[\sigma] =\mathring D[\Sigma^+[\sigma]]$; we use the standard notation of a bar to denote the closure of a set and ring for the interior. See \Fig{fig:lightsheets}.

\begin{figure}[t]
\centering
\includegraphics[width=.8\textwidth,page=1]{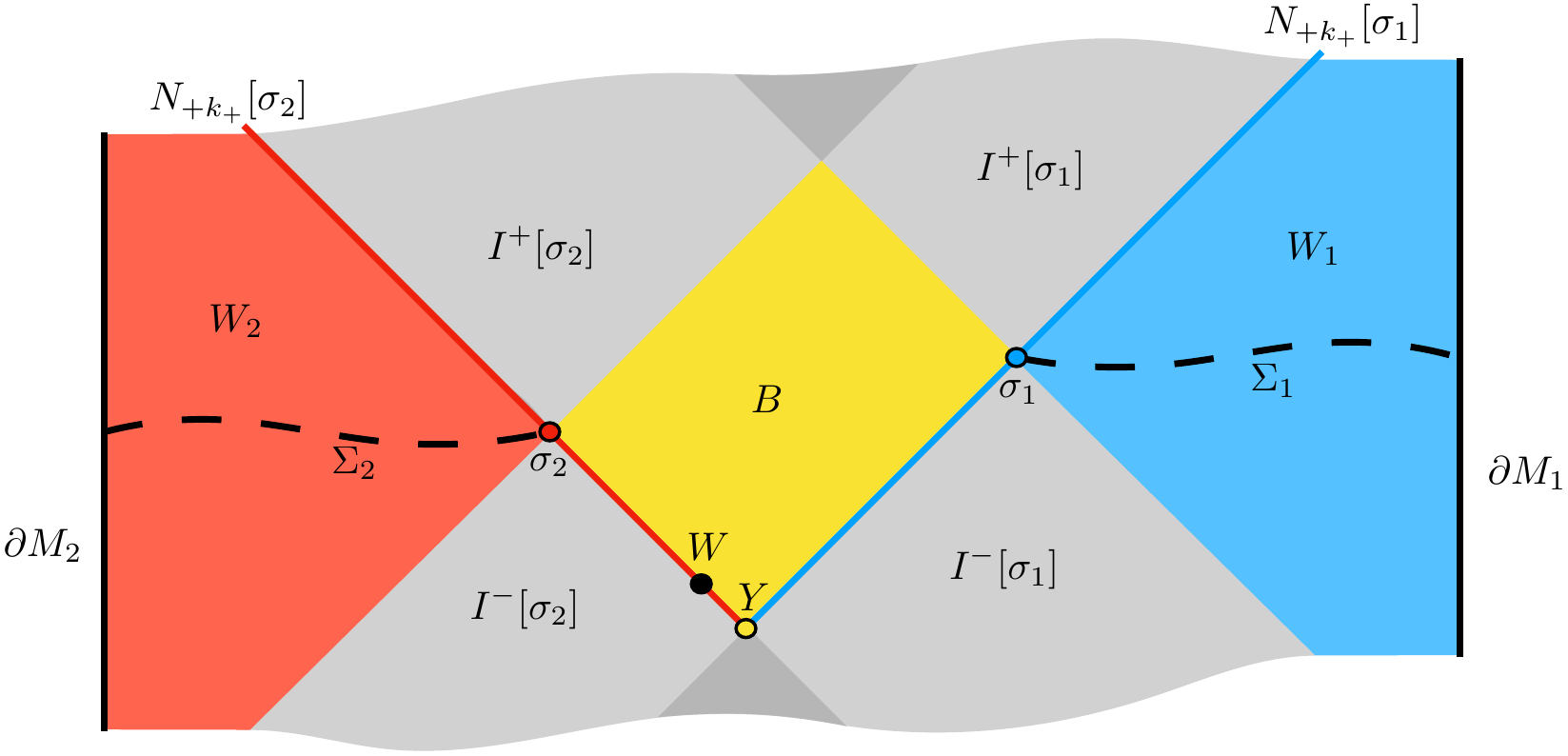}
\caption{Conformal diagram depicting the construction of the spacetime $M$ containing both $\sigma_1$ and $\sigma_2$ and the identification of an extremal surface $W$ of maximum area, as described in text.\label{fig:bridge}}
\end{figure}

\begin{figure}[t]
\centering
\includegraphics[width=.75\textwidth,page=1]{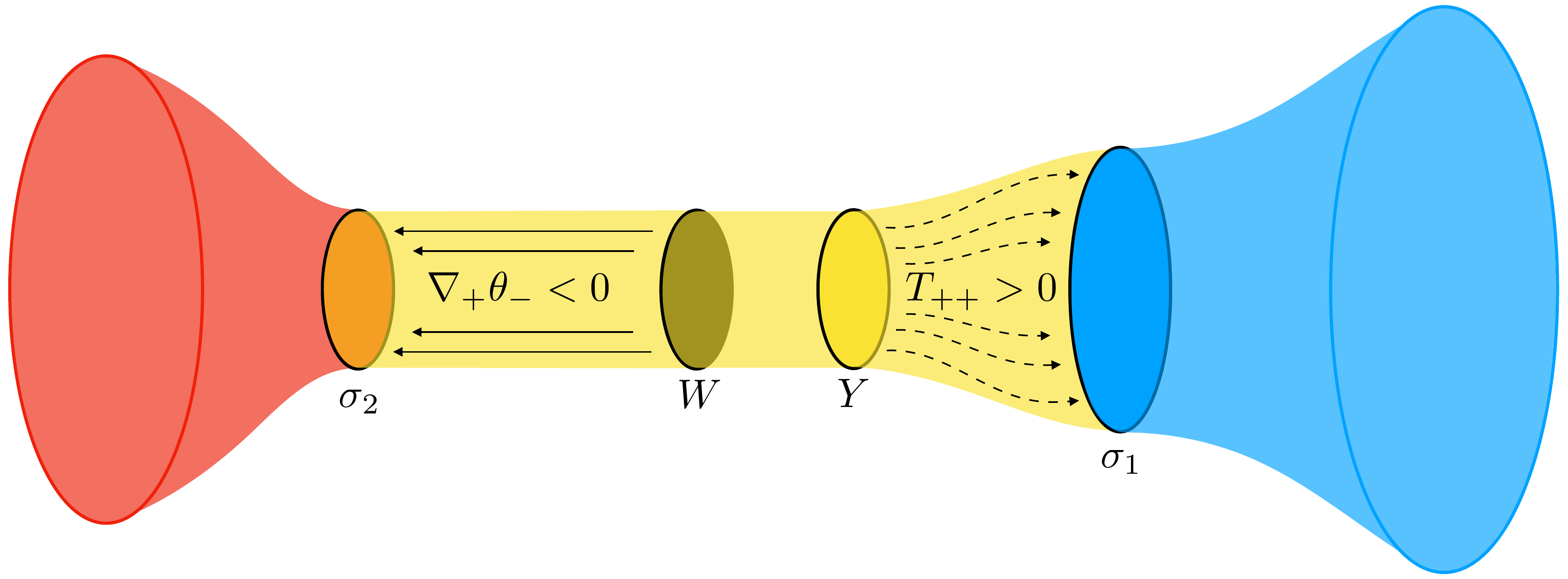}
\caption{Depiction of the spatial slice (for the $A[\sigma_1]>A[\sigma_2]$ case) comprised of $\Sigma_1$, $\Sigma_2$ and the past boundary of $B$. Nonzero $T_{++}$ on $N_{-k_+}[\sigma_1]$ from $Y$ to $\sigma_1$ is identified with dashed arrows, and $\nabla_k \theta_-$ on  $N_{-k_+}[\sigma_2]$ from $W$ to $\sigma_2$ is identified with solid arrows (in the direction of decreasing $\theta_-$).
While we have colored the disks inside the black circles defining $\sigma_{1,2}$, $W$, and $Y$ in the lower figure for ease of labeling, the surfaces themselves are in fact only the boundary of these disks. \label{fig:bridge2}}
\end{figure}

Given $W_1 \equiv W[\sigma_1]\subset M_1$ and $W_2 \equiv W[\sigma_2] \subset M_2$, we wish to construct a spacetime containing $W_1$ and $W_2$.
We thus assemble a spacetime $M$ that is composed of seven parts: $W_1$, $W_2$, $\overline I^\pm[\sigma_1]$, $\overline I^\pm[\sigma_2]$, and a final ``bridge'' region that we will write as $B\equiv \mathring D[\Sigma^-_1[\sigma_1]]\cap \mathring D[\Sigma^-_2[\sigma_2]]$, denoting the interior of the intersection of the two inner wedges of $\sigma_1$ and $\sigma_2$.
See Figs.~\ref{fig:bridge} and \ref{fig:bridge2} for an illustration.
Throughout, we will require purity of the state holographically dual to $M$; geometrically, this requires that $M$ be inextendible and in particular that it contain both sides of any apparent horizon.

How can we be sure that such a spacetime exists? The characteristic initial data formalism (CIDM) guarantees the existence of a spacetime in the domain of dependence of some Cauchy surface $\Sigma$ defined as a piecewise union of light sheets, provided that one supplies data on $\Sigma$ satisfying the constraint equations~\cite{Rendall:2000,Brady:1995na,ChoquetBruhat:2010ih,Luk:2011vf,Chrusciel:2012ap,Chrusciel:2012xf,Chrusciel:2014lha}. Along the $k_+$ light sheet, the constraint equations are:
\begin{equation}
\begin{aligned}
\nabla_{+} \theta_{+} &= -\frac{1}{D-2}\theta_+^2 - \varsigma^2 - 8\pi G T_{++} & [\text{Raychaudhuri}] \\
q_\mu^{\;\;\nu} {\cal L}_+ \omega_\nu &= -\theta_+ \omega_\mu + \frac{D-3}{D-2} {\cal D}_\mu \theta_+ - ({\cal D}\cdot\varsigma)_\mu + 8\pi G T_{\mu +} & [\text{Damour-Navier-Stokes}] \\
\nabla_+ \theta_- &= -\frac{1}{2}{\cal R} - \theta_+ \theta_- +\omega^2 + {\cal D}\cdot \omega + 8\pi G T_{+-}, & [\text{cross-focusing}]
\end{aligned}
\end{equation}
where the replacement of any Lorentz index with $\pm$ denotes its contraction with $k_\pm$, ${\cal L}_{+}$ is the Lie derivative along $k_+$, $q_{\mu\nu}$ is the induced metric, ${\cal D}_\mu = q_\mu^{\;\;\nu}\nabla_\nu$, $\varsigma$ is the shear describing tidal deformations of the light sheet, $\omega$ is the twist describing rotation of the light sheet, and ${\cal R}$ is the intrinsic Ricci scalar, all evaluated on a constant affine parameter slice.
The null congruences are normalized such that $k_+ \cdot k_- = -1$.\footnote{Note that the CIDM is simply a way of reframing the Einstein equations as an initial-value problem, with the null surface of the light sheet playing the role of the Cauchy surface; specifying $T_{\mu\nu}$, $\varsigma$, and ${\cal R}$, one has a set of three coupled first-order equations for $\theta_+$, $\theta_-$, and $\omega$ along each affine parameter, for which a solution generically exists.}

Given our two apparent horizons $\sigma_1$ and $\sigma_2$, the spacetime $M$ is essentially a special type of wormhole, containing $\sigma_1$ and $\sigma_2$ as mouths.
This wormhole will contain some HRT surface $X$~\cite{Hubeny:2007xt,Dong:2016hjy}, computable via the maximin prescription~\cite{Wall:2012uf} (and we will see explicitly how to construct it later in this section).
That is, $X$ is a surface for which both orthogonal null congruences have vanishing expansion and for which there exists some Cauchy slice on which $X$ is the minimal cross section.
In AdS/CFT, the area of this surface computes the entanglement between the state on $\partial M_1$ and its purification:
$S[\rho_1] = -{\rm tr} \rho_1 \log \rho_1 = A[X]/4G\hbar$.

Given an apparent horizon, a useful coarse-grained holographic information theoretic measure is the {\it outer entropy}, 
\begin{equation}
S^{(\rm outer)}[\sigma] = \max_{\rho}\left. S[\rho]\right|_{W[\sigma]\text{ fixed}},\label{eq:outerS}
\end{equation}
where the maximization is taken over all reduced density matrices $\rho$ defined on the AdS boundary dual to $W[\sigma]$ for which the geometry on $W[\sigma]$ is fixed.
Geometrically, given a purification of $\rho$ describing an inextendible spacetime containing $W[\sigma]$, $S^{({\rm outer})}[\sigma]$ computes the maximum area of the HRT surface dividing $W[\sigma]$ from the other asymptotic AdS region(s).
If $\sigma$ is marginally trapped (i.e., for an apparent horizon), Ref.~\cite{Engelhardt_2018} showed that $S^{({\rm outer})}[\sigma] = A[\sigma]/4G\hbar$, while a more general algorithm for computing the outer entropy for normal surfaces was constructed in Refs.~\cite{Nomura_2018,Bousso_2019}.

As a refinement of the outer entropy, we can consider a constrained maximization problem.
Rather than maximize the HRT area over {\it all} purificiations of the wedge $W_1$ as in Eq.~\eqref{eq:outerS}, we can instead maximize subject to the constraint that the new spacetime one constructs for the purification also contain some other apparent horizon $\sigma_2$ and its associated outer wedge $W_2$.
We can regard the set of all such purifications---specifically, the boundary state dual to the ``extended'' wedge $D[W_1 \cup B]$---as a {\it mesostate} of the black hole with apparent horizon $\sigma_1$: given the exterior $W_1$, we have specified a subset of the physical properties of the microstate, namely, that it contains the marginally-trapped surface $\sigma_2$. Note that we are considering everywhere-geometric purifications in which $\sigma_1$ remains a horizon, in contrast to Ref.~\cite{Bao_2017}, which remains agnostic as to the nature of the region behind the horizon; that is, in the purifications we consider here, the horizon is not replaced by stringy degrees of freedom.
Maximizing the area of $X$ over all such mesostates defines a {\it subsystem outer entropy}:
\begin{equation}
S^{({\rm outer})}[\sigma_1;\sigma_2]=\max_{\rho_1}\left. S[\rho_1] \right|_{W_1,W_2\text{ fixed}} = \max_M\left. \frac{A[X]}{4G\hbar}\right|_{W_1,W_2\text{ fixed}}.
\end{equation}
 This reduces to the conventional definition of the outer entropy when we take the special purification where $\sigma_{1}=\sigma_{2}$ and $W_{1}$ and $W_{2}$ are two copies of the same wedge.
 We expect that $S^{({\rm outer})}[\sigma_1;\sigma_2] \leq A[\sigma_1]/4G\hbar$, since we are holding more fixed in our definition.
 By symmetry of the definition under swapping labels $1\leftrightarrow 2$, we should have
 \begin{equation}
 S^{({\rm outer})}[\sigma_1;\sigma_2] \leq \frac{\min \{ A[\sigma_1],A[\sigma_2]\}}{4G\hbar}.\label{eq:ineqS}
 \end{equation}

Let us compute the subsystem outer entropy. 
First, as in Ref.~\cite{Nomura_2018}, we prove that $X$ must lie within $\overline B$.
We proceed by contradiction, supposing that $X$ is not contained in the closure of $B$.
Define $\Sigma$ as a Cauchy slice of $M$ on which $X$ is the minimal cross section, which exists since $X$ is a HRT surface. Let $\xi^{\pm}=\Sigma\cap N_{\pm k_{\mp}}[\sigma_{1}]$. Either $\xi^{+}$ or $\xi^{-}$ will be nonempty, and we define this surface as $\xi$.
By the Raychaudhuri equation and null energy condition that $T_{++},T_{--}\geq 0$ (NEC), $A[\xi]\leq A[\sigma_{1}]$. 
Since $X$ is the minimal cross section on $\Sigma$, we must have that $A[\rho]\geq A[X]$. 
Hence, $A[X]\leq A[\sigma_{1}]$. 
Now, $N_{\pm k_{\pm}}[X]$ will intersect $\Sigma_{1}$---the partial Cauchy surface with respect to which $\sigma_1$ is outermost---on some surface $\zeta\neq\sigma_{1}$. 
By the Raychaudhuri equation and NEC, $A[\zeta]\leq A[X]$. 
By the ``outermost'' hypothesis for $\sigma_{1}$, $A[\zeta]>A[\sigma_{1}]$. 
Then $A[X]>A[\sigma_{1}]$, contradicting our result that $A[X]\leq A[\sigma_{1}]$. 
Thus, $X\subset \overline B$.

Let us now define $\mu_{1,2}^{\pm}=\Sigma\cap N_{\pm k_{+}}[\sigma_{1,2}]$. 
Either $\mu_{1}^{+}$ or $\mu_{1}^{-}$ will be nonempty (which we designate $\mu_1$), and similarly either $\mu_{2}^+$ or $\mu_2^-$ will be nonempty (which designate $\mu_2$). 
Since $\theta_{+}[\sigma_{i}] = 0$, the NEC and Raychaudhuri equation imply that cross sections of $N_{\pm k_{+}}[\sigma_{i}]$ have decreasing area as we move away from $\sigma_{i}$, so $A[\mu_{1}]\leq A[\sigma_{1}]$ and $A[\mu_{2}]\leq A[\sigma_{2}]$. 
But since $\Sigma$ is the surface on which $X$ is minimal, it follows that $A[X]\leq A[\mu_{1}]$ and $A[X]\leq A[\mu_{2}]$. 
Hence, $A[X] \leq \min \{A[\sigma_1],A[\sigma_2] \}$ for any HRT surface $X \subset M$, and so we find that Eq.~\eqref{eq:ineqS} is indeed satisfied.

We now argue that the inequality in Eq.~\eqref{eq:ineqS} is in fact saturated.
To do so, we demonstrate the opposite weak inequality, $\max A[X] \geq \min \{A[\sigma_1],A[\sigma_2] \}$, by exhibiting a particular spacetime in which $A[X] = \min \{A[\sigma_1],A[\sigma_2] \}$.
The construction is closely related to that in Refs.~\cite{Engelhardt_2018,Nomura_2018,Bousso_2019}, making use of the CIDM.
Suppose, without loss of generality, that $A[\sigma_2] < A[\sigma_1]$ (if the opposite inequality obtains, swap $1\leftrightarrow 2$ in the rest of the construction, while if they are equal, the problem reduces to that of Ref.~\cite{Engelhardt_2018} as noted above).
On $N_{-k_+}[\sigma_2]$, we choose the energy-momentum tensor to satisfy $T_{++} = T_{+\mu} = 0$, as well as $\varsigma = 0$, while keeping $T_{+-}$, $\omega_\mu$, and ${\cal R}$ constant.\footnote{The shear need not vanish on $\sigma_2$ itself, since we can arrange a discontinuity in the shear via a gravitational wave sourcing a shock in the Weyl tensor~\cite{Wald}.}
The Damour-Navier-Stokes equation thus becomes trivial, while the Raychaudhuri equation guarantees that $\theta_+$ vanishes along $N_{-k_+}[\sigma_2]$, so that all slices of this light sheet have the same area.
The light sheet $N_{-k_+}[\sigma_2]$ therefore cannot become singular, and it meets $N_{-k_+}[\sigma_1]$ on some surface $Y$.
The cross-focusing equation (along with the generic condition) implies that $\theta_-$ increases at constant affine rate as we flow back along the light sheet, at some point reaching zero on a surface $W$.\footnote{Recall that $\theta_+$ must vanish on $W$, too, since we have arranged $T_{++} = \varsigma = 0$ along $N_{-k_+}[\sigma_2]$ from $\sigma_2$ to $Y$.}

From $W$, we can define the two future-directed orthogonal null congruences, which we will call $\ell_1$ and $\ell_2$, with $\ell_i$ directed toward $\sigma_i$.
In general, it will not be the case that $\ell_i = k_+[\sigma_i]$, since, e.g., $k_+[\sigma_2]$ is orthogonal to slices of $N_{-k_+}[\sigma_2]$ of constant affine parameter, while $W$ itself may not be such a slice, since different generators of the light sheet can reach $\theta_- = 0$ at different affine parameter values, and similarly for the $\sigma_1$ light sheet.
Now, $\theta_-$ on $N_{-k_+}[\sigma_2]$ and $\theta_{\ell_2}$ on $W$ are related by a second-order equation, and one can show that the existence of a $\theta_- = 0$ surface on the light sheet guarantees the existence of a bona fide extremal surface, where $\theta_{\ell_1} = 0$, on the light sheet~\cite{Engelhardt:2018kcs}.
However, a simpler approach is to use the gauge freedom in the normalization of $k_+$ to rescale the affine parameter generator-by-generator along the light sheet so that it corresponds to $\ell_2$, guaranteeing that $W$ indeed is simultaneously marginally trapped and marginally antitrapped~\cite{Bousso_2019}.
Similarly, we can rescale the normalization of $k_+$ on $\sigma_1$ so that $Y$ is also a surface of constant affine parameter on the $N_{-k_+}[\sigma_1]$ light sheet.

We must have consistent, single-valued initial data on the past boundary of $B$. Since $A[Y] = A[\sigma_2] < A[\sigma_1]$ by our choices of data on $N_{-k_+}[\sigma_2]$, the Raychaudhuri equation and NEC along $N_{-k_+}[\sigma_1]$ then imply that---if we fix the shear to zero---we must have strictly positive $T_{++}$ somewhere along the $\sigma_1$ light sheet in order to supply the area difference. (We assume here that any nonlocal self-intersections of null geodesics~\cite{Akers:2017nrr} along $N_{-k_+}[\sigma_1]$ that happen to fall on $Y$ cover a set of measure zero, thereby avoiding double counting any finite area.
This condition holds generically, since for it to fail would require a fine-tuned alignment of $Y$ with such a region of self-intersection, and we can avoid such an alignment by introducing shear along $N_{-k_+}[\sigma_1]$.)
Thus, the Raychaudhuri equation implies that $Y$ has positive null expansion in the direction of $\sigma_1$, which means that $Y \neq W$, and in particular (given the cross-focusing equation and our choices of data along $N_{-k_+}[\sigma_2]$) that $Y$ is to the past of $W$ along $N_{-k_+}[\sigma_2]$. 
Thus, $W$ is contained in the past boundary of $B$.

We have shown that $W$ is extremal (i.e., both null expansions vanish) and---consistent with what we have proven about the HRT surface---that $W\subset \overline B$. 
To guarantee that $W$ is an HRT surface, it remains to exhibit a Cauchy surface on which $W$ is a minimal cross section.
By the Raychaudhuri equation, NEC, and the outermost conditions, our discussion above implies that such a surface is given by the union of the segment of the $N_{-k_+}[\sigma_2]$ light sheet from $\sigma_2$ to $Y$ and the segment of the $N_{-k_+}[\sigma_1]$ light sheet from $Y$ to $\sigma_1$, along with the partial Cauchy slices $\Sigma_1$ and $\Sigma_2$ with respect to which $\sigma_1$ and $\sigma_2$ satisfy the ``outermost'' criterion.
We therefore find that $W=X$, the HRT surface, and we have $\max A[X] \geq A[\sigma_2]$ when $A[\sigma_2]<A[\sigma_1]$.
We have thus proven our desired result:
\begin{equation}
S^{({\rm outer})}[\sigma_1;\sigma_2] = \frac{\min \{ A[\sigma_1],A[\sigma_2]\}}{4G\hbar}.
\end{equation}

\section{Black Hole Mesostates}\label{sec:meso}

Let us now return to our microstate discussion of Sec.~\ref{sec:micro}, with an eye toward generalization.
When $A$ is the entire boundary CFT, the $\rho_{A,i}$ were chosen to be pure, in other words, states of the same entanglement entropy, namely zero. When $A$ was taken to be smaller than the entire boundary CFT, these states were allowed to become mixed, but we still required that all the states have the same, now nonzero, value for their von~Neumann entropy. The calculations for the Holevo information continued working in this case.

There is another way of generating a new distribution from the microstates for which the $\rho_{A,i}$ all have the same nonzero entanglement entropy, even when $A$ is the entire boundary region of the CFT. As noted in Sec.~\ref{sec:intro}, this can be accomplished by ``block renormalizing'' the microstates together, grouping them in subsums $\bar\rho_{A,(n,m)}=\sum_{i=n}^{n+m} p_i \rho_{A,i}$, which we will write as $\bar \rho_{A,j}$. We choose the subsets of microstates such that the entanglement entropy of all of the $\bar\rho_{A,j}$ is the same and $\rho_A=\sum_j \bar p_j \bar\rho_{A,j}$. As introduced in Sec.~\ref{sec:intro}, we denote the sum over the black hole microstates implied by these $\bar\rho_{A,j}$ black hole mesostates---a term introduced by Sorkin~\cite{sorkin1999large}---and the geometries defined analogously to the black hole microstate geometries as black hole mesostate geometries.
Of course, there is enormous freedom in the choice among all possible block renormalizations, and thus a correspondingly large family of mesostates.
We choose to group the microstates into particular mesostates such that each of the individual mesostates $\bar \rho_{A,j}$ describes, in the bulk dual, a geometric region larger than the wedge of $\rho_A$.
This grouping is an intermediate coarse/fine-graining between the thermofield double and the black hole microstate cosmologies described in Ref.~\cite{Cooper_2019}. As both of these geometries are everywhere smooth, it is well motivated that the mesostate geometry should similarly be everywhere smooth, at least in the region between $A$ and its HRT surface.
That is, the mesostates are those described by the extended wedge of Sec.~\ref{sec:subsystem}, for some surface $\sigma_2$ (see \Fig{fig:bridge2} for an example).

One can then ask the Holevo distinguishability questions again in this context. First, we note that the computations for $S(\rho_A)$ are precisely the same as in Sec.~\ref{sec:micro}. The difference arises in the computation of the $S(\bar \rho_{A,j})$. Now, unlike in Sec.~\ref{sec:micro}, there is a homology constraint sourced by the Shannon entropy of the mesostate geometries, i.e., the horizon of the new surface $\sigma_2$. When $A[\sigma_2]<A[\sigma_1]$ (where $\sigma_1$ is the event horizon of the original black hole), the new homology constraint is weaker, and we have $S(\bar \rho_j) = A[\sigma_2]/4G\hbar$ (which we write as $S_{{\rm BH}_2}$), where $\bar \rho_j$ is the state of entire boundary CFT in the mesostate. This means that when $A$ itself is the entire boundary CFT, we have $\chi=S_{\rm BH}-S_{\rm BH_2}$, i.e., $(A[\sigma_1]-A[\sigma_2])/4G\hbar$, which is smaller than $S_{\rm BH}$. One can ask what happens in the case where $A[\sigma_1]\leq A[\sigma_2]$. In this situation, as shown in \Sec{sec:subsystem}, the HRT prescription would settle on $\sigma_1$, as it is the smaller of the two surfaces. From the perspective of the partitioning of the microstate ensemble, this would correspond to the trivial partition, as the entropy cannot be larger than the total number of microstates. Moreover, this eliminates the possibility of surfaces with areas larger than the smaller of $\sigma_1$ and $\sigma_2$ from having an entropic interpretation as a function of $\rho_A$.
Thus, when $A$ comprises the entire connected component(s) of the CFT homologous to and outside of the original horizon, the Holevo information $\chi$ associated with the mesostate is directly related to the subsystem outer entropy of Sec.~\ref{sec:subsystem}:
\begin{equation}
\chi = S_{\rm BH} - S^{\rm (outer)}[\rm BH_1;BH_2],\label{eq:holevoidentity}
\end{equation}
where we write $\rm BH_{1,2}$ for $\sigma_{1,2}$. This result is exactly what we wanted to find. Because entropies are logarithmic quantities, the difference $A[\sigma_1]-A[\sigma_2]$ corresponds to the modding out of the Hilbert space associated with the block renormalization scheme used above from the full Hilbert space. Indeed, the Shannon entropy of the mesostate distribution is precisely $S_{\rm BH}-S_{\rm BH_2}$. Therefore, once again when we have access to the entire boundary we have perfect distinguishability over the mesostates. A larger amount of distinguishability here would have been contradictory, as it would have implied the ability to distinguish between states that have been block renormalized with observables too coarse for that distinguishability task.

\begin{figure}[t]
\centering
\includegraphics[width=.25\textwidth]{holodisc4.pdf} \quad \includegraphics[width=.25\textwidth]{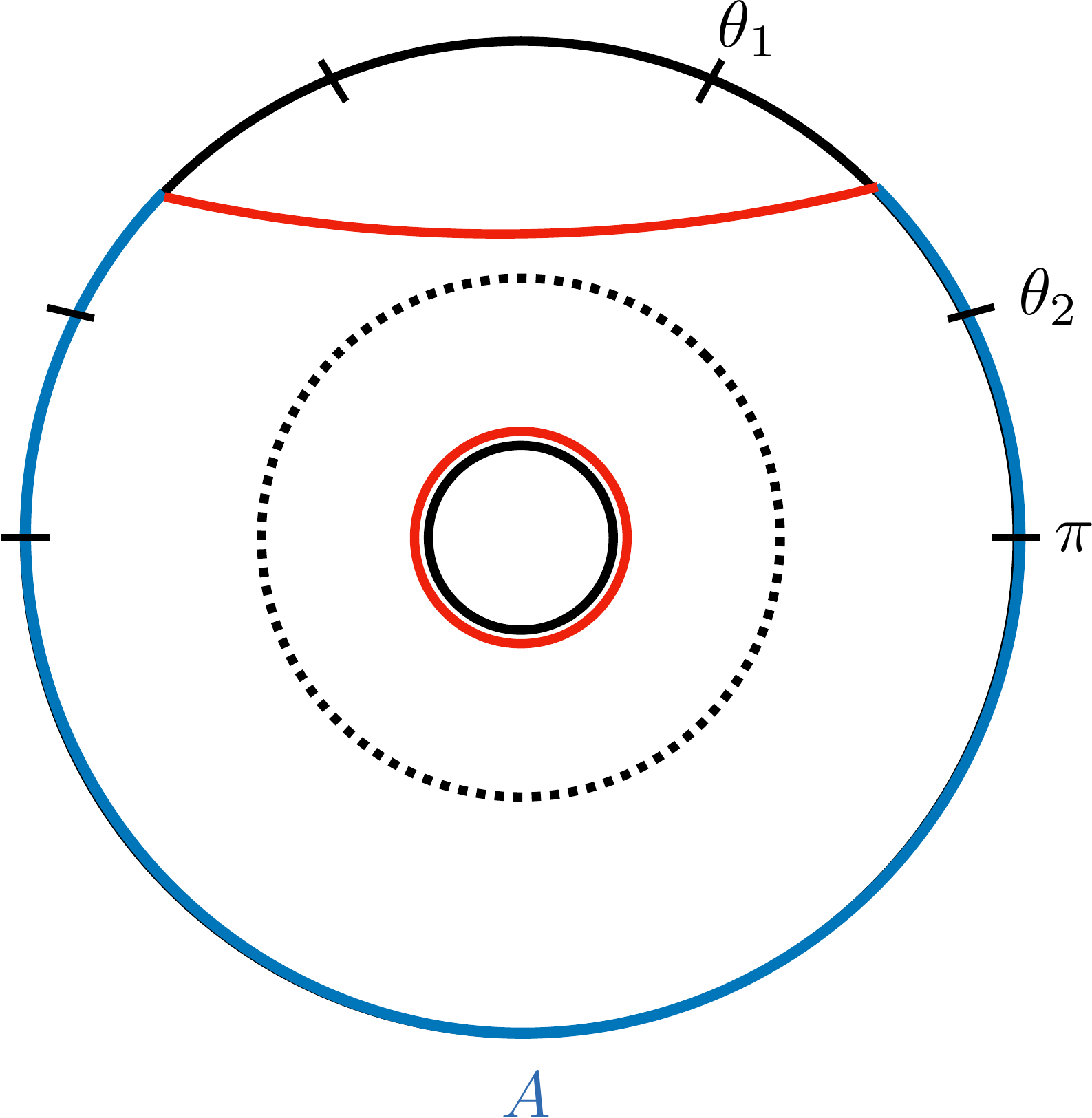} \quad \includegraphics[width=.25\textwidth]{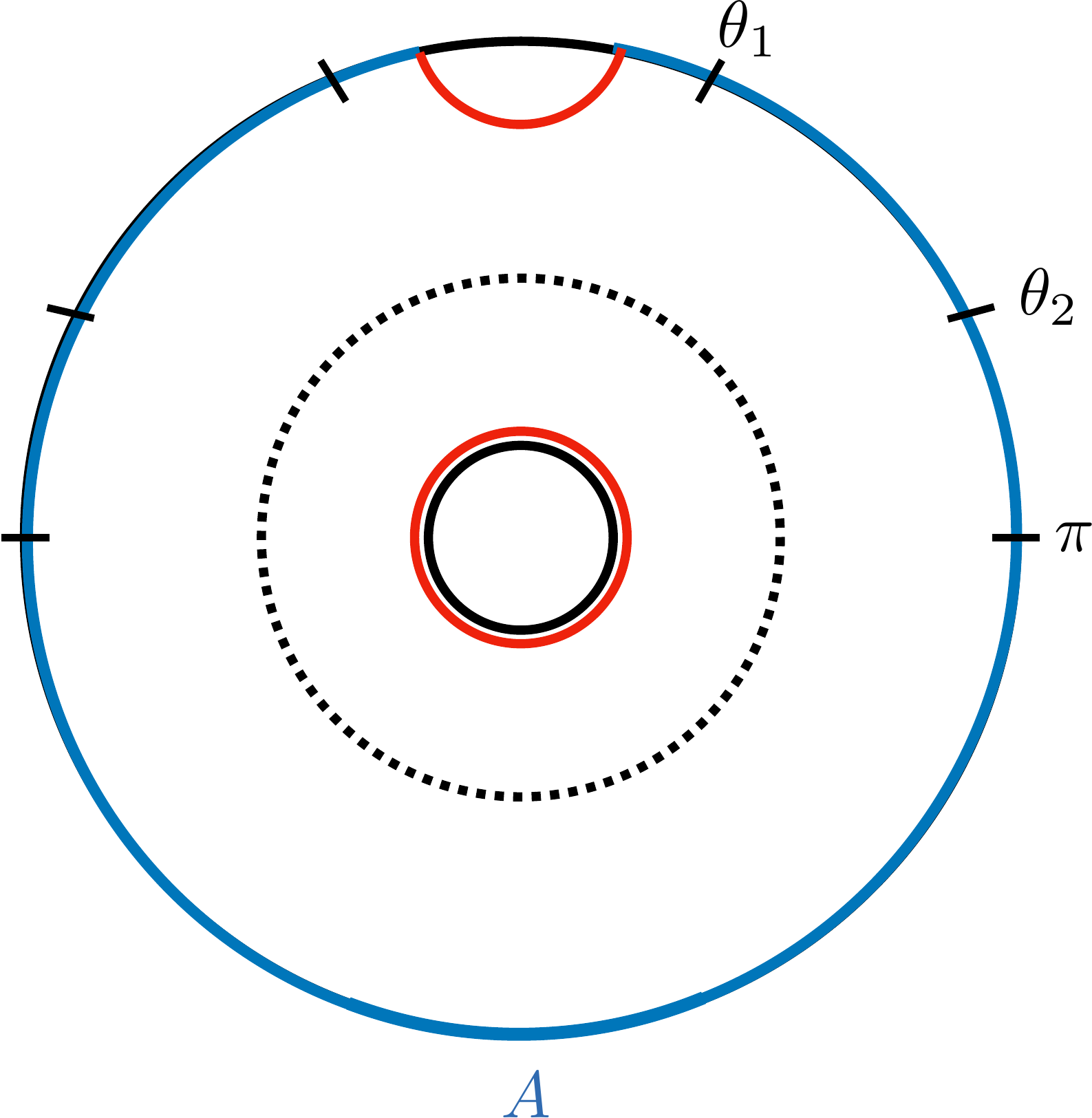}
\caption{\label{fig:meso} The minimal surface of the mesostate differs from the black hole because the geodesic generically can enter past $r_1$ and into the internal region. At $\theta_A=\theta_2$, the homology condition requires a thermal contribution from the smaller horizon with radius $r_2$ (solid black circle). As the size of region $A$ increases beyond $\theta_2$, the minimal surfaces of the black hole and mesostate are the same except for the mismatch of the horizon sizes.}
\end{figure}

Having established this check, the remaining Holevo information calculations from Ref.~\cite{Bao_2017} can all be repeated here. In particular, we expect that the Holevo information should remain zero before $A$ is sufficiently large (half the system size in ${\rm AdS}_3/{\rm CFT}_2$), will rise with shallow slope until the RT phase transition of the mesostate geometry is reached, then rise with steeper slope until the RT phase transition of the full geometry is reached, and thereafter saturate at $S_{\rm BH}-S_{\rm BH_2}$.
The reason that the slope generically increases after the mesostate phase transition is that the Holevo information is defined as a difference of two functions, and the function being subtracted---namely, the entropy of the partially-purified state---has smaller overall entanglement entropy, which consequently grows more slowly with size of $A$ than the original entropy.
The typical behaviors of the minimal surfaces for the various regimes of the black hole, microstate, and mesostate are shown in Figs.~\ref{fig:BH},~\ref{fig:micro}, and \ref{fig:meso}, respectively.

\begin{figure}[t]
\centering
\includegraphics[width=.45\textwidth]{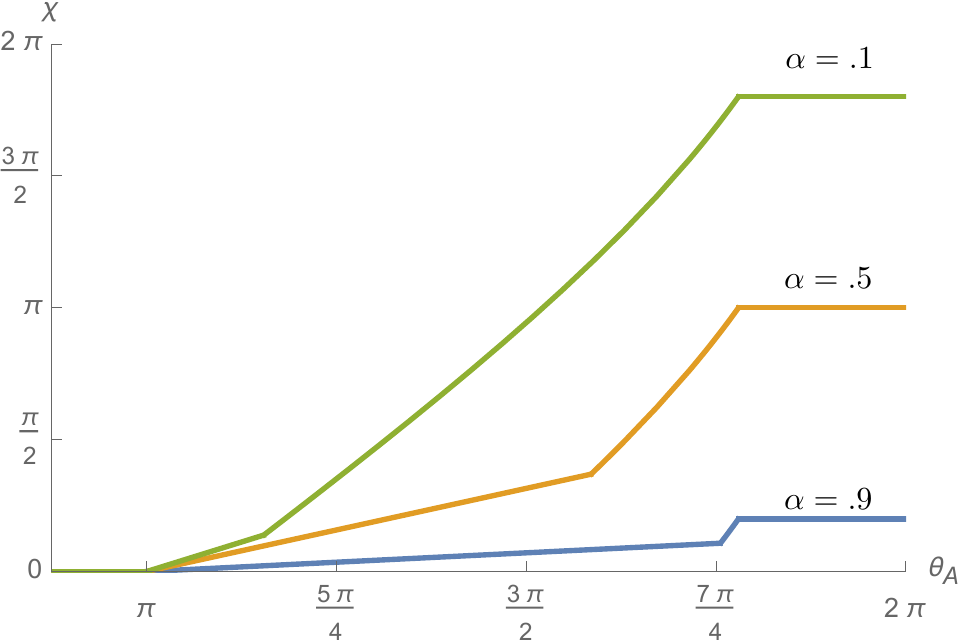} \quad
\includegraphics[width=.45\textwidth]{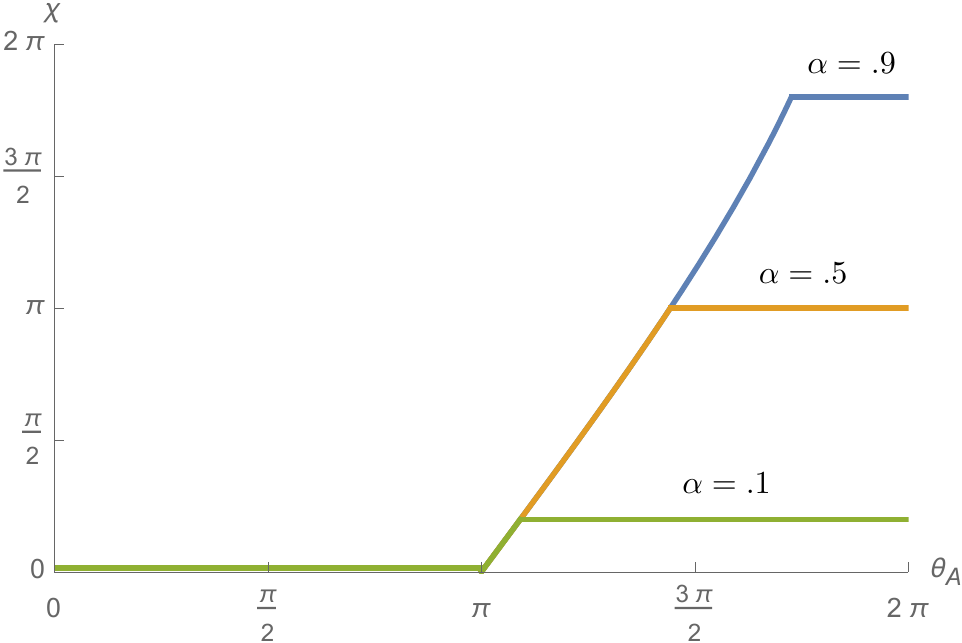} 
\caption{\label{fig:chi} Generic behavior of $\chi_{\rm BH|meso}(\theta_A)$ (left) and $\chi_{\rm meso|micro}(\theta_A)$ (right) for $\alpha=0.1$ in green, $0.5$ in orange, and $0.9$ in blue.}
\end{figure}

From the properties of the minimal surfaces of these geometries, we can describe the general behavior of the Holevo information. Because there are two different sums we consider---a sum over some mircostates to form a mesostate and a sum over mesostates to form a black hole state---there are two different Holevo informations depending on which ensemble is considered:
\begin{equation}\label{eq:chis}
    \begin{split}
        \chi_{\rm BH|meso}&\equiv S_{\rm BH}-S^{\rm (outer)}[\rm BH_1;BH_2] \\
        \chi_{\rm meso|micro}&\equiv S^{\rm (outer)}[\rm BH_1;BH_2]-S_{\rm micro}.
    \end{split}
\end{equation}
Choosing $r_2=\alpha r_1$, where $0\leq \alpha \leq 1$, we obtain the curves for the Holevo information depicted in \Fig{fig:chi}.

We expect that these information theoretic observations are well complemented by the holographic dual description of the completion of a black hole geometry to another of a different, smaller temperature. We anticipate that the additional stress-energy needed to engineer this wormhole geometry is the matter needed to partially purify the larger black hole with horizon $\sigma_1$. Furthermore, because the second black hole $\sigma_2$ is simply an auxiliary system to facilitate calculation of the mesostate entanglement entropies, without loss of generality it can be taken to be homologous to the horizon of the primary black hole of interest, thus eliding potential topology-matching concerns.
   
\section{Example: Multiboundary Wormholes}\label{sec:multiboundary}

We will now proceed with an explicit calculation of the Holevo information using a particular class of geometries of multiboundary wormholes~\cite{Krasnov:2000zq,Krasnov:2003ye,Skenderis:2009ju,Balasubramanian:2014hda,Maxfield_2015}. These spacetimes are formed by choosing a collection of boundary regions of ${\rm AdS}_3$ and identifying various internal geodesics that bound the resulting homology region. (While the three-dimensional analogue of Birkhoff's theorem sharply curtails the geometries that can be constructed consistent with axial symmetry~\cite{AyonBeato:2004if}, it is possible to construct arbitrarily complicated wormhole geometries via general gluings of minimal surfaces.) Importantly for our purposes, one can independently choose the size of the minimal throats homologous to each boundary as well as the moduli that control the topology of the internal region. From this approach, we can explicitly construct our micro- and macrostates as well as the fully purified black hole state.

\begin{figure}[t]
\centering
\includegraphics[width=.3\textwidth]{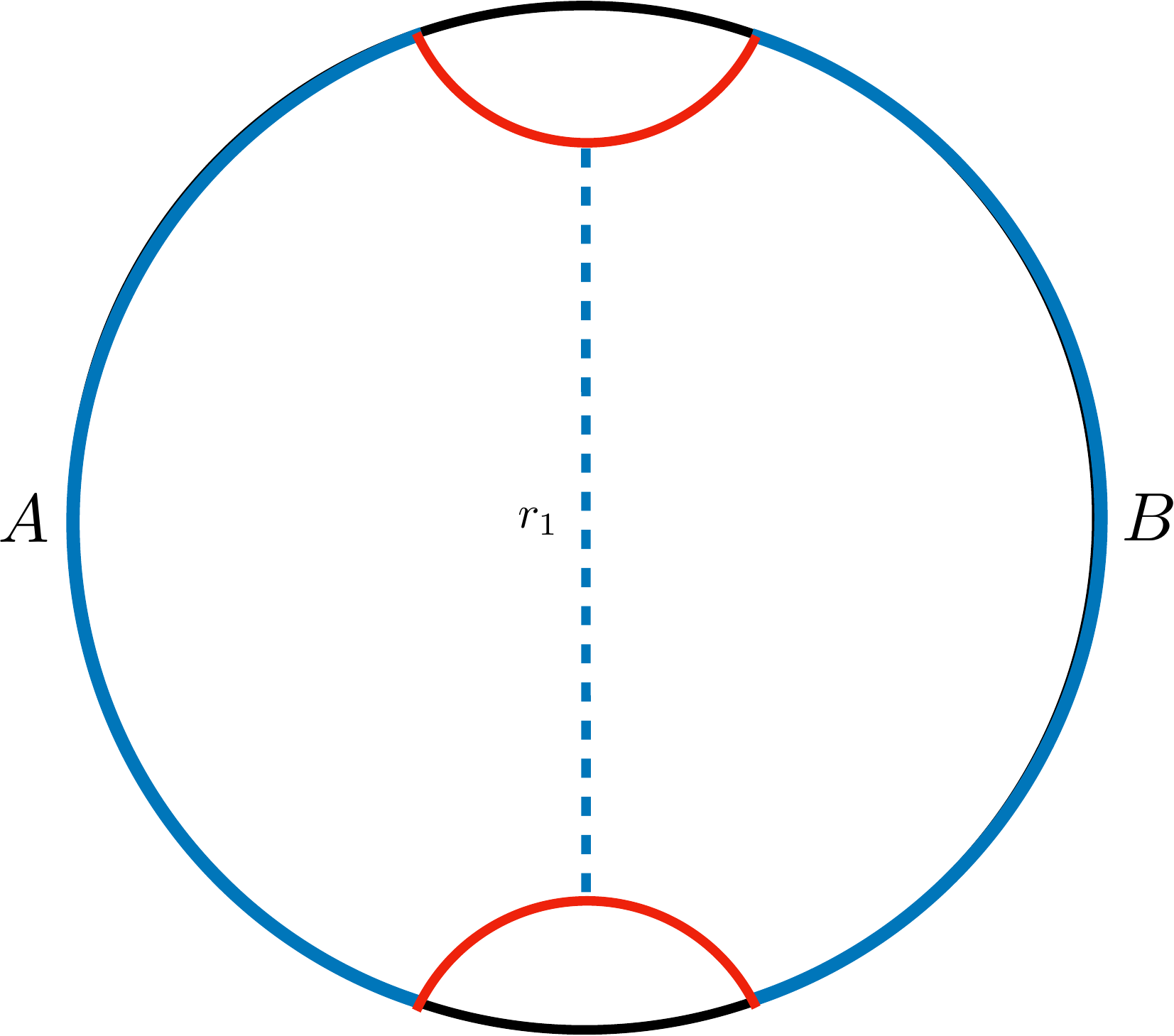} \quad \raisebox{1.6em}{\includegraphics[width=.5\textwidth]{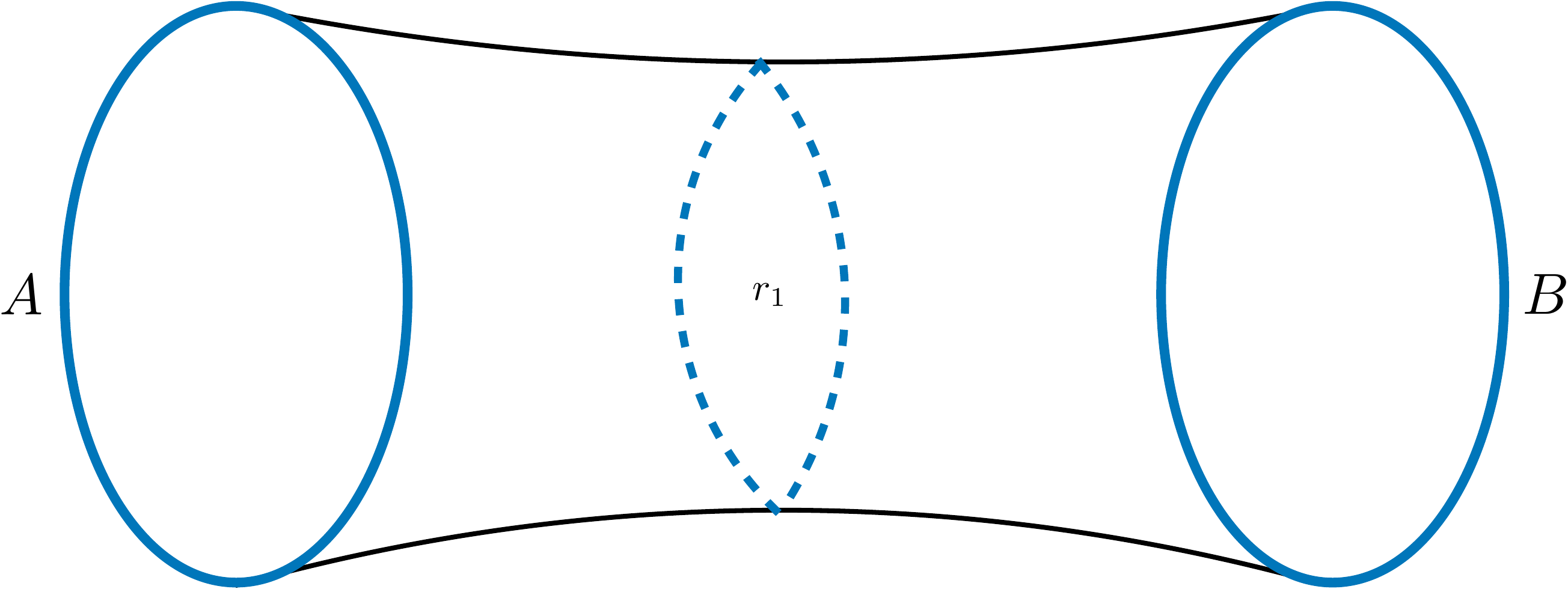}} 
\caption{\label{fig:MB_BH}Construction of the black hole geometry. The red geodesics are identified to form the fully purified thermofield double state with horizon size $r_1$. }
\end{figure}

The black hole state is formed by starting with two boundary regions $A$ and $B$ and considering their homology region. We take the entanglement wedge cross section---the length of the minimal surface homologous to $A$ relative to the entanglement wedge $m(AB)$---to have size $r_1$. We then perform an identification of the two disconnected boundaries of $m(AB)$. The resulting geometry is a thermofield double state, a two-boundary wormhole with throat size $r_1$; see Fig.~\ref{fig:MB_BH}.

The resulting metric can be written as that of a static BTZ black hole with horizon size $r_1\equiv r$, from which we can calculate the entanglement entropy $S(A)$ as a function of region size. We parameterize $A$ by the angle $\theta_A$ of the subtended region. Because of the homology condition, the entanglement entropy will undergo a phase transition at the critical angle
\begin{equation}
\theta_{1}=\frac{\ell}{r}\log\left(\frac{1+e^{2\pi r/\ell}}{2}\right),
\end{equation}
where the thermal contribution $2\pi r$  appears. (We write $\ell$ for the AdS length.) The entanglement entropy~\cite{Hubeny_2013} is given by
\begin{equation}
S_{\rm BH}(\theta_A)=
    \begin{cases}
    \frac{\ell}{2G\hbar}\log\left[\frac{2\epsilon}{r}\sinh(\frac{r}{2\ell}\theta_A)\right] & \theta_A< \theta_1 \\
      \frac{\ell}{2G\hbar}\log\left[\frac{2\epsilon}{r}\sinh(\frac{r}{2\ell}(2\pi-\theta_A))\right]+ \frac{\pi r}{2G\hbar} & \theta_1\leq \theta_A,
    \end{cases}
\end{equation}
where $\epsilon$ is the UV cutoff on the boundary.
   
\begin{figure}[t]
\centering
\includegraphics[width=.4\textwidth]{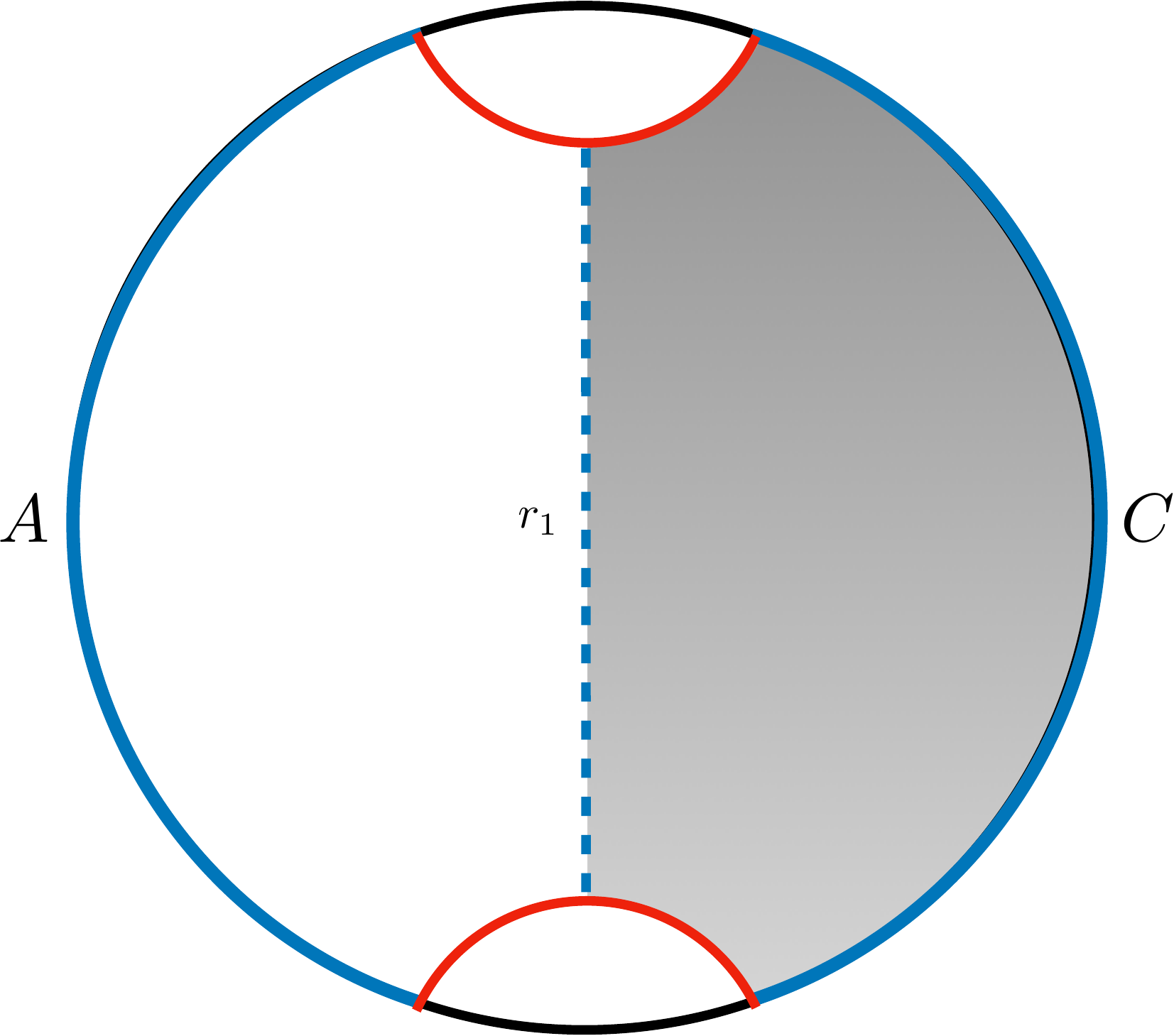}
\caption{\label{fig:MB_micro}A microstate geometry. The gray region represents bulk degrees of freedom. }
\end{figure}

A microstate is by definition a pure state with the same geometry up to $r_1$, beyond which is unknown and generically nongeometric. However, any degrees of freedom past $r_1$, which we call $C$, can under the surface-state correspondence~\cite{Miyaji_2015} be thought of as internal bulk degrees of freedom associated with the original black hole (see Fig.~\ref{fig:MB_micro}). As a result, the entanglement entropy should be calculated as the minimal surface \emph{relative} to the bulk degrees of freedom. It can never be advantageous---from an area-minimization perspective---for a boundary-anchored geodesic to end on one of these internal boundaries, since such a geodesic can never contribute favorably to satisfying the homology constraint. As such, the entanglement entropy is given by the same minimal geodesics as the black hole geometry, but \emph{without} any thermal contributions:
\begin{equation}
S_{\rm micro}(\theta_A)=
    \begin{cases}
     \frac{\ell}{2G\hbar}\log\left[\frac{2\epsilon}{r}\sinh(\frac{r}{2\ell}\theta_A)\right] & \theta_A< \pi \\
     \frac{\ell}{2G\hbar}\log\left[\frac{2\epsilon}{r}\sinh(\frac{r}{2\ell}(2\pi-\theta_A))\right] & \pi\leq \theta_A.
    \end{cases}
\end{equation}
The phase transition occurs at $\theta_{A}=\pi$, where the minimal surface jumps from on side of the horizon to the other.

\begin{figure}[t]
\centering
\raisebox{1.31em}{\includegraphics[width=.4\textwidth]{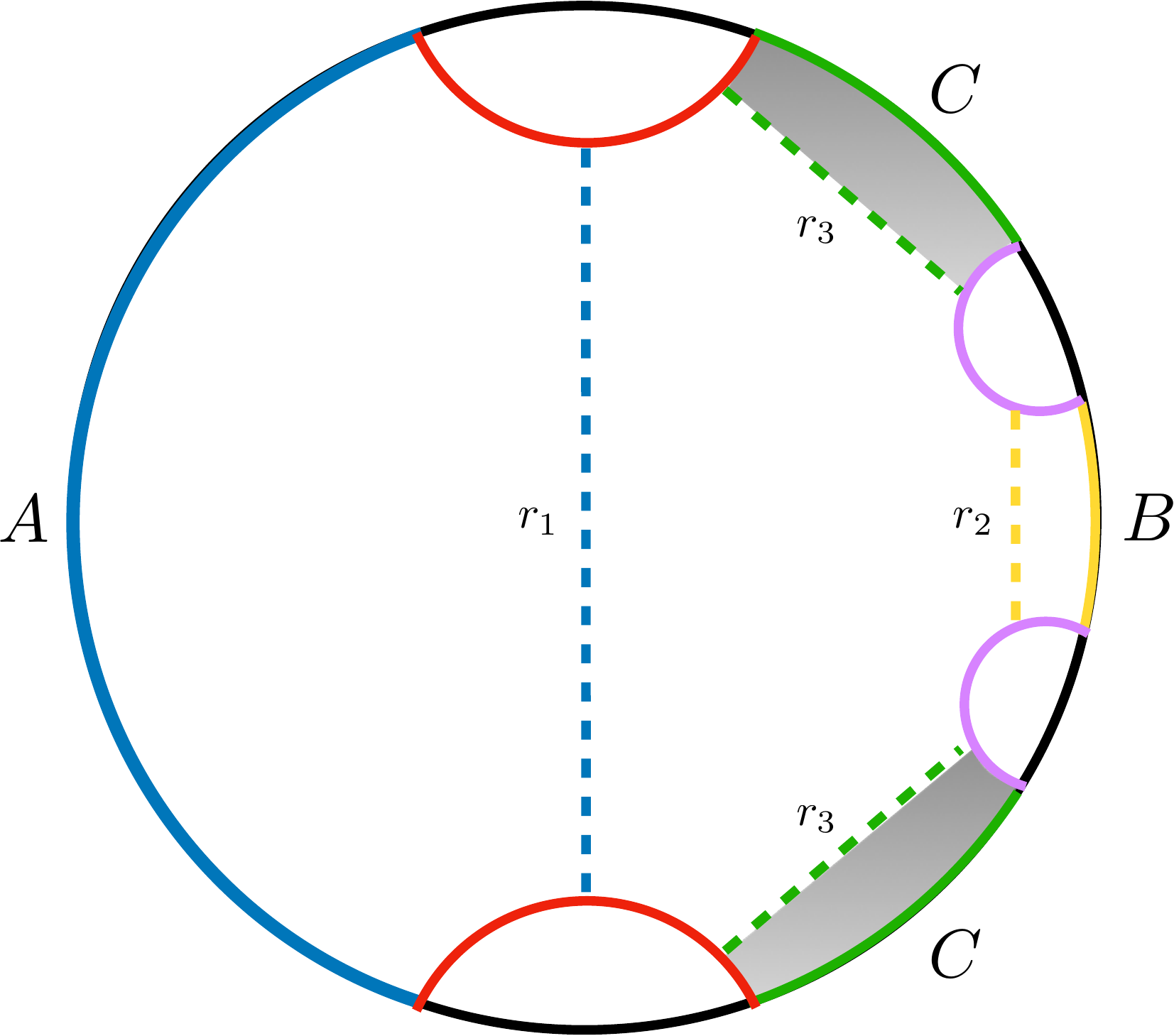}} \quad \includegraphics[width=.46\textwidth]{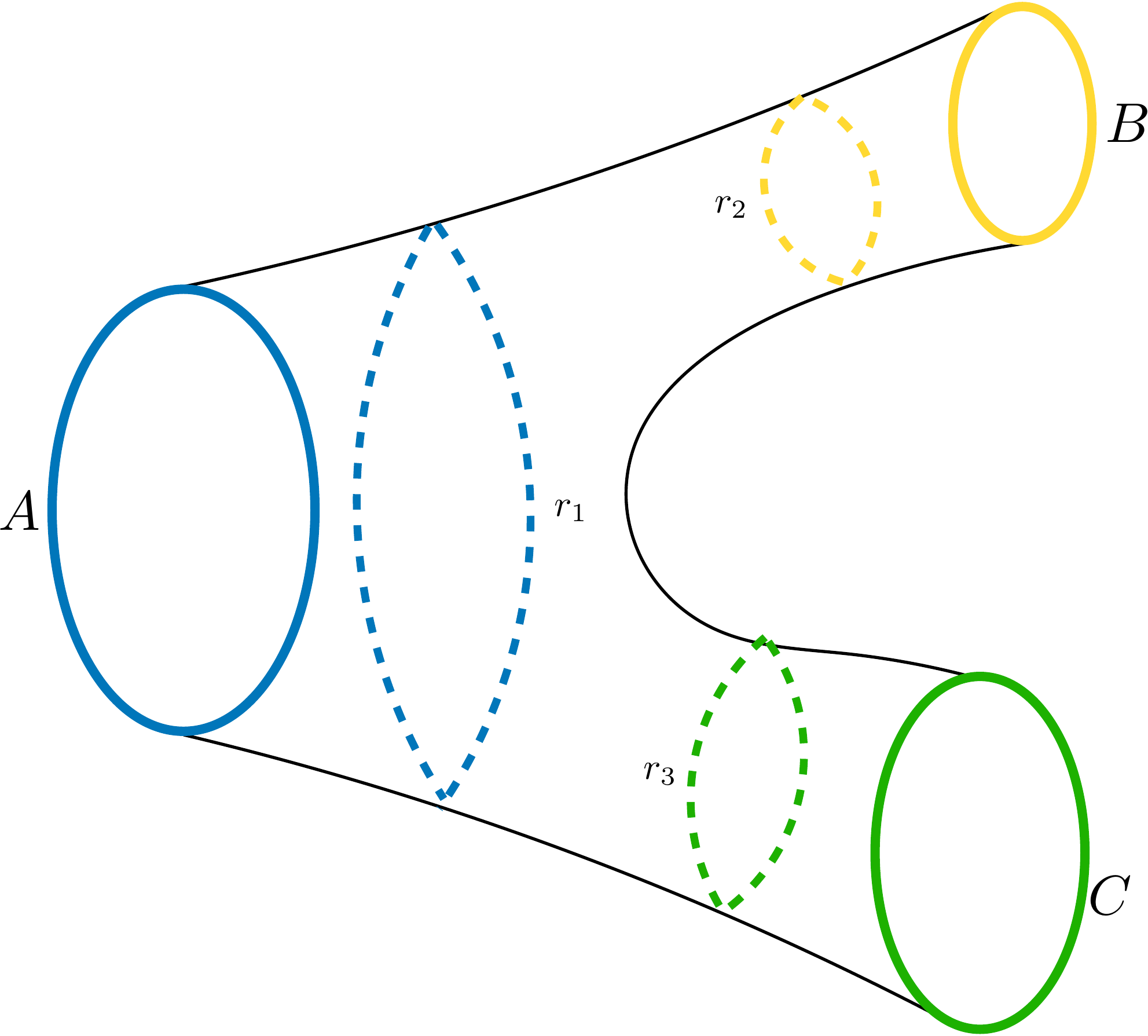} 
\caption{\label{fig:MB_meso}The construction of a mesostate as a three-boundary wormhole geometry. The two red geodesics are identified, as is the purple pair. A partial sum over microstates partially purifies the geometry, causing the internal region between $r_1$, $r_2,$ and $r_3$ to be geometric.}
\end{figure}

Finally, we construct the mesostate geometry as a three-boundary wormhole with boundaries $A,B,C$ (see Fig.~\ref{fig:MB_meso}). Our original boundary $A$ is given a throat of size $r_1\equiv r$ so that the geometry exterior to the horizon is identical to the black hole. The area of horizon $\sigma_2$, which is homologous to $B$, is written as $r_2=\alpha r_1$, where $0\leq \alpha \leq 1$ so that it is strictly smaller than $r_1$. The boundary $C$ should once again be viewed as a collection of bulk degrees of freedom; we take $r_3$, the circumference of the throat homologous to $C$, to an arbitrary value so long as $r_2+r_3\geq r_1$, so that $\sigma_1$ is minimal. We also require that the geometry is in the fully-connected phase. These choices are made in order to mirror the setup of \Sec{sec:subsystem}.

To calculate $S(A)$, we must again consider minimal surfaces homologous to $A$ relative to $C$. In this situation, this simply means that the minimal surface should create a homology region separating $A$ from $B$, but it does not necessarily need to separate $A$ from $C$. The result will be the same as the black hole, but with the thermal contribution and critical angle calculated with respect to the smaller horizon, $r_2=\alpha r$, and in the BTZ geometry with horizon length $r_1=r$. (The reason for this is that, in this construction built out of pure ${\rm AdS}_3$, the minimal geodesics do not penetrate the region between $r_1$ and $r_2$ \cite{Maxfield_2015}.) The resulting entropy is
\begin{equation}
S_{\rm meso}(\theta_A)=
    \begin{cases}
      \frac{\ell}{2G\hbar}\log\left[\frac{2\epsilon}{r}\sinh(\frac{r}{2\ell}\theta_A)\right] & \theta_A< \theta_2 \\
     \frac{\ell}{2G\hbar}\log\left[\frac{2\epsilon}{r}\sinh(\frac{r}{2\ell}(2\pi-\theta_A))\right]+ \frac{\pi \alpha r}{2G\hbar} & \theta_2\leq \theta_A,
    \end{cases}
\end{equation}
with critical angle given by
\begin{equation}
\theta_2 = \frac{\ell}{r}\log\left(\frac{1+e^{\pi r(\alpha+1)/\ell}}{1+e^{\pi r(\alpha-1)/\ell}}\right).
\end{equation}

\begin{figure}[t]
\centering
\includegraphics[width=.5\textwidth]{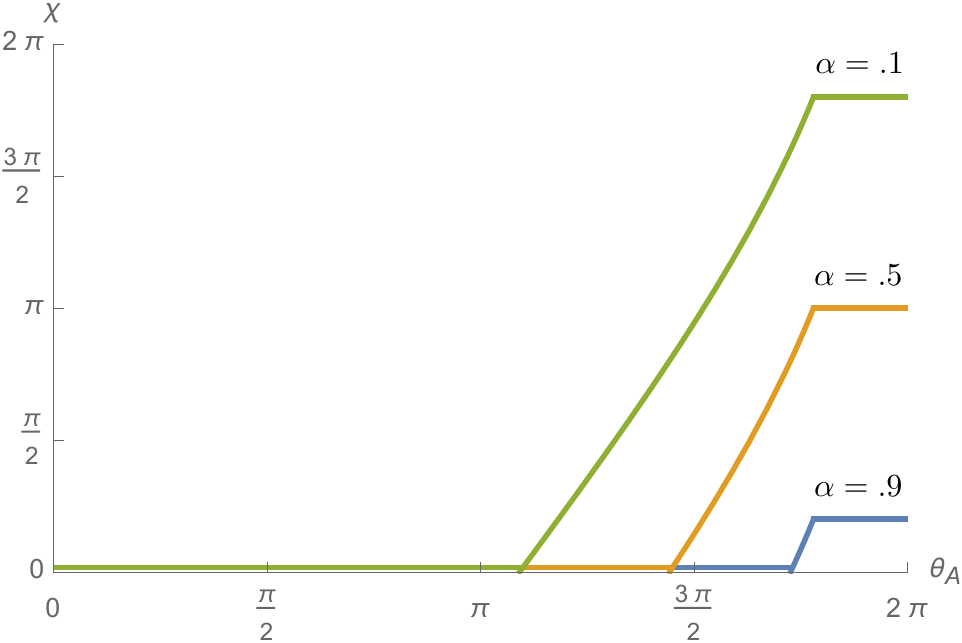} 
\caption{\label{fig:chi_BH_meso} The Holevo information $\chi_{\rm BH|meso}(\theta_A)$ for $\alpha=0.1$ in green, $0.5$ in orange and, $0.9$ in blue. Here we have taken $4G\hbar=\ell=1$ and $r=\ell$. Note that, in this example constructed from identifications of empty ${\rm AdS}_3$, the minimal geodesics of the mesostate geometry do not extend past $r_1$ and into the interior region. As a result, this Holevo information remains zero until the phase transition at $\theta_2$.\label{fig:chi1}}
\end{figure}

\begin{figure}[H]
\centering
\includegraphics[width=.5\textwidth]{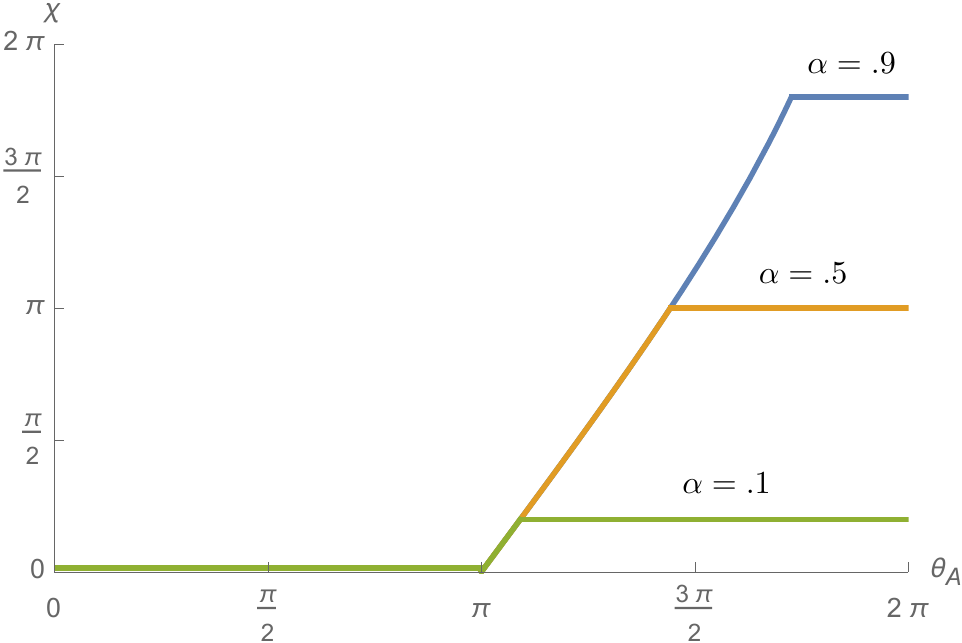} 
\caption{\label{fig:chi_meso_micro} The Holevo information $\chi_{\rm meso|micro}(\theta_A)$ for $\alpha=0.1$ in green, $0.5$ in orange and, $0.9$ in blue. As in \Fig{fig:chi1}, we have taken $4G\hbar=\ell=1$ and $r=\ell$. \label{fig:chi2}}
\end{figure}

From the three geometries, we can subtract the entropies to get the Holevo information between the black hole and mesostates as well as between the meso- and microstates (see \Eq{eq:chis}):
\begin{equation}
\chi_{\rm BH|meso}(\theta_A)=
\begin{cases} 
     0 & \theta_A\leq \theta_2 \\
       \frac{\ell}{2G\hbar}\log\left[\frac{\sinh(\frac{r}{2\ell}\theta_A)}{\sinh(\frac{r}{2\ell}(2\pi-\theta_A))}\right]- \frac{\pi \alpha r}{2G\hbar} & \theta_2<  \theta_A \leq \theta_1 \\
       \frac{\pi  r}{2G\hbar}(1- \alpha) & \theta_1 < \theta_A, 
   \end{cases}
   \end{equation}
shown in \Fig{fig:chi1}, and 
  \begin{equation}
\chi_{\rm meso|micro}(\theta_A)=
\begin{cases} 
     0 & \theta_A\leq \pi \\
      \frac{\ell}{2G\hbar}\log\left[\frac{\sinh(\frac{r}{2\ell}\theta_A)}{\sinh(\frac{r}{2\ell}(2\pi-\theta_A))}\right] & \pi<\theta_A\leq \theta_2 \\
       \frac{\pi \alpha r}{2G\hbar} & \theta_2 < \theta_A ,
   \end{cases}
   \end{equation} 
shown in \Fig{fig:chi2}.
\vspace{3mm}

\vspace{3mm}

\section{Discussion}\label{sec:discussion}

In this paper, we have investigated the holographic interpretation of Holevo information in the context of distinguishability of black hole mesostates, partial purifications of black hole states that are formed by coarse-grainings of microstates.

In \Sec{sec:subsystem}, we generalized the concept of holographic outer entropy---itself a useful coarse-grained entropic quantity encoded in the geometry---to the case in which one conditions on some information about the interior (i.e., subsystem information): specifically, given a marginally-trapped surface $\sigma_1$, we showed that the maximal HRT surface interior to $\sigma_1$, consistent with holding the exterior geometry fixed and conditioned on the existence of another marginally-trapped surface $\sigma_2$ in the interior, is equal to the area of the smaller of $\sigma_{1,2}$ (times $1/4G\hbar$).

In \Sec{sec:meso}, we considered the geometric interpretation of Holevo information, showing this quantity to be closely connected with the subsystem outer entropy of geometric black hole mesostates, in which the coarse-graining is performed in such a way that the entropy of the subsystem is dual to the area of the interior surface.
We constructed an explicit example demonstrating these concepts through the use of multiboundary wormholes built from ${\rm AdS}_3$ in \Sec{sec:multiboundary}, holographically computing the Holevo information curves and examining their phase transitions.

We believe that studying models of black hole microstates, in conjunction with the previously discussed models of black hole mesostates as wormhole geometries, will lead to significant progress on the problem of black hole statistical mechanics.
In the following discussion, we highlight two potential avenues of future study: a wormhole-based organization of black hole microstates and connections of our mesostate constructions to the bit threads formalism.

\paragraph{A Wormhole Approach to Microstate Counting}
Consider a wormhole with three throats, of areas $A$, $A/2$, and $A/2$. To each of the $A/2$ throats, glue via Ref.~\cite{Engelhardt_2018} another three-throated wormhole of areas $A/2$, $A/4$, and $A/4$. Iterating in this fashion, we can construct a wormhole with binary tree-like connectivity, until we reach branches with Planck-area cross sections (or some large constant multiple thereof, if we wish to preserve a semiclassical spacetime).
The fact that the total area remains the same at each level in this construction suggests the possibility of a unitary implementation of the state transformation from layer to layer without the need for isometries (that is, the quantum circuit implementing the corresponding tensor network is not changing the entanglement entropy between the input and the output, and thus is permitted to be unitary).
\begin{figure}[t]
\centering
\includegraphics[width=.44\textwidth]{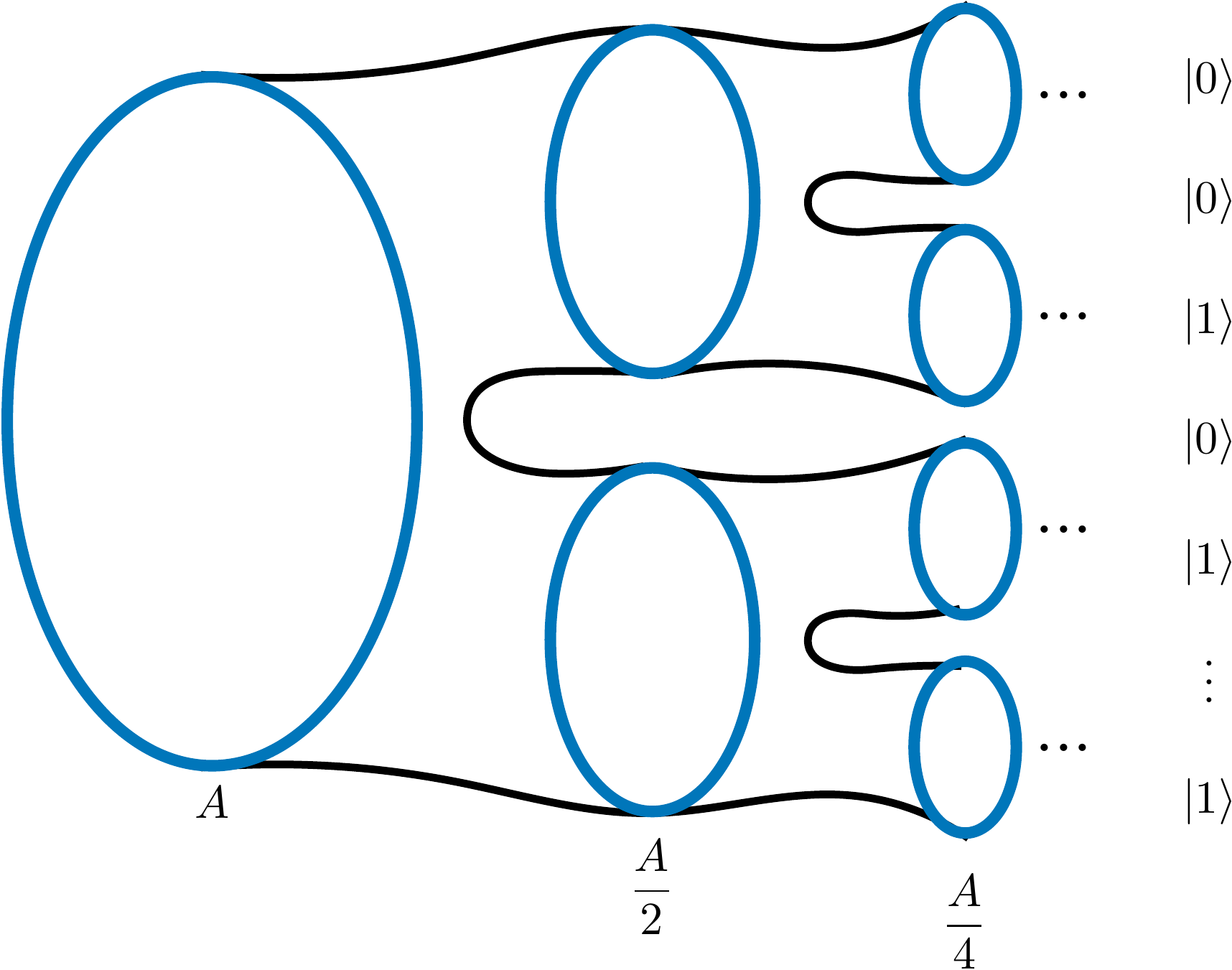} 
\caption{\label{fig:MS_count} A wormhole geometry that branches to Planck-sized horizons, which can be thought of as being purified by individual qubit states. The state defined on these qubits can be mapped to a unique microstate, with the branch structure defining some list of microstate properties.}
\end{figure}
One can view the terminus of each of these branches as dual to different CFT boundaries, analogous to Ref.~\cite{penington2020entanglement}, or in terms of a qubit on an end-of-the-world brane \`a la Ref.~\cite{Cooper_2019}. See \Fig{fig:MS_count}.

The state of these qubits will correspond to the individual black hole microstates. While a typical black hole microstate will {\it not} itself possess this tree-like structure, these branching geometries and the qubit purifications thereof are unitarily related to the black hole microstate geometries in question.
One can view the tree-like structure as a dichotomous key, where the location of a given terminal branch within the tree corresponds to a binary string encoding various properties of the microstate. 
(Encoding the full degrees of the freedom of the quantum state in this way would additionally require encoding phase information between the qubits at the ends of different branches.)
This model is attractive because the different qubit purifications give a microscopic accounting of the black hole microstates in a way that enables the asking of statistical mechanics questions, such as studying the fluctuation-dissipation relations \cite{kubo1966fluctuation} and the Jarzynski inequalities \cite{jarzynski2011equalities} in the context of black holes.

\paragraph{Bit threads}

\begin{figure}[t]
\centering
\includegraphics[width=.44\textwidth]{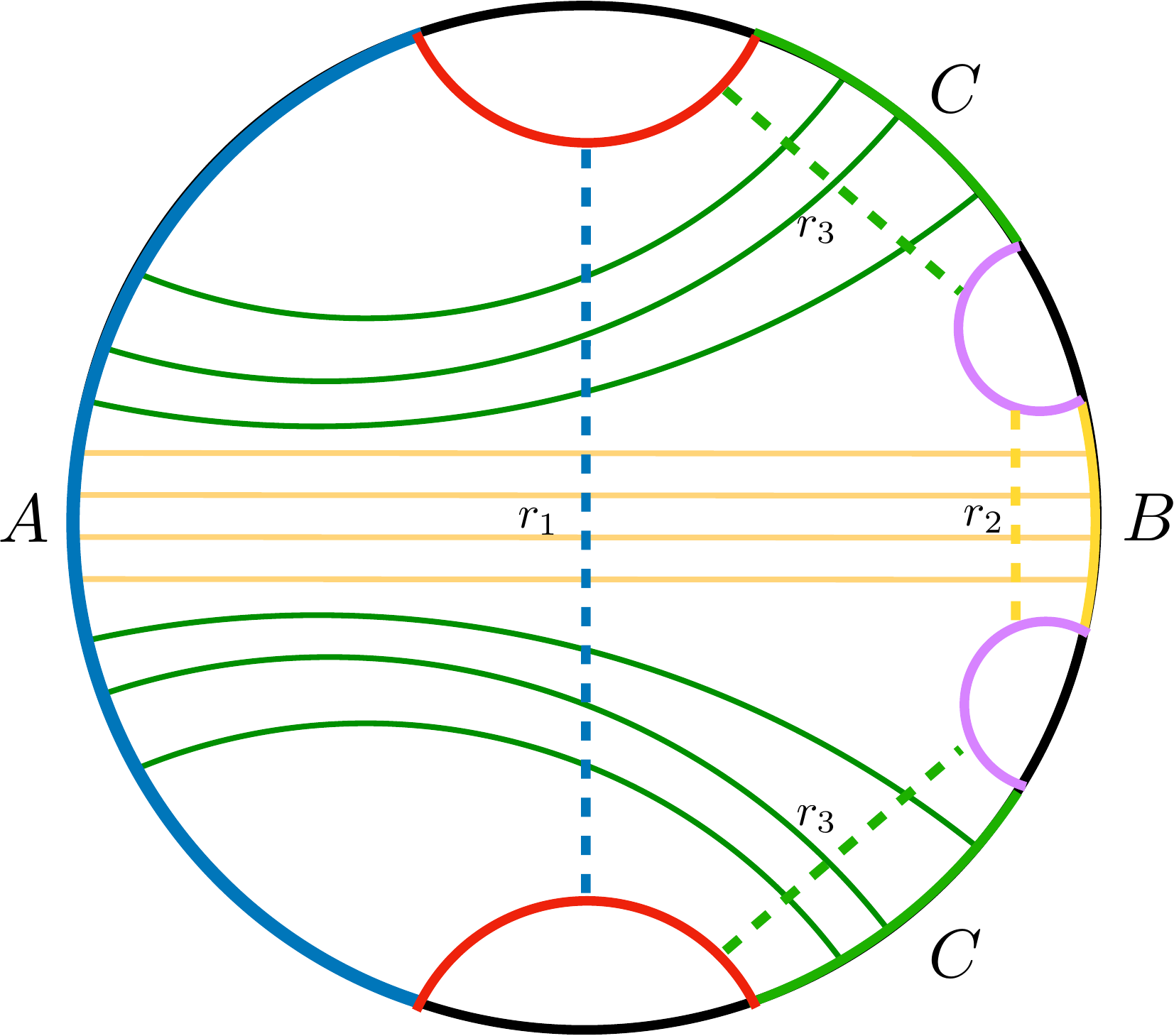} 
\caption{\label{fig:BT_Holvo} A maximal bit thread configuration on $A$ confined to the homology region of $ABC$ can be chosen to saturate on both $r_1$ and $r_2$. This configuration can be lifted to the identified multiboundary wormhole geometry i.e. the mesostate. The number of threads that connect $A$ and $C,$ shown in green, calculate the Holevo information $\chi_{\rm BH|meso}$.}
\end{figure}
The bit thread formalism \cite{Freedman_2016,Headrick_2018,Harper_2019} allows for entanglement entropy to be calculated as a maximal collection of threads that connect two points on the boundary CFT. In this language, the Holevo information can be constructed as follows: Starting from a mesostate geometry as defined previously one sends as many threads from $A$ as possible with the added condition that the threads remain inside the homology region of $ABC$. Since the threads are taken to possess a finite thickness, the maximum number of them that can be sent is given by the area of the minimal surface $\sigma_1$. Each of these threads must either end on $B$ or $C$. We can opt to choose our configuration to additionally satisfy the condition that $\sigma_2$ is also saturated. This essentially amounts to a gauging of the solution to require that as many threads reach $B$ as possible. The remaining threads necessarily connect $A$ and $C$. Now, the total number of threads from $A$ is $2\pi r_1$ while the number of these threads that reach $B$ is $2\pi r_2$. This means that the remaining threads connecting $A$ and $C$ is given precisely by the Holevo information (see Fig.~\ref{fig:BT_Holvo}).
An interesting direction of future study would be to further investigate the question of black hole mesostates in a bit threads context.

\section*{Acknowledgments}
We would like to thank Nathan Benjamin, William Donnelly, Matt Headrick, and Brian Swingle for useful discussions.  N.B. is supported by the Department of Energy under grant number DE-SC0019380 and by the Computational Science Initiative at Brookhaven National Laboratory. 
The work of J.H. is supported by the Simons Foundation through \emph{It from Qubit: Simons Collaboration on Quantum Fields, Gravity, and Information}.
G.N.R. is supported at the Kavli Institute for Theoretical Physics by the Simons Foundation (Grant No.~216179) and the National Science Foundation
(Grant No.~NSF PHY-1748958) and at the University of California, Santa Barbara by the Fundamental Physics Fellowship.

\bibliographystyle{utphys-modified}
\bibliography{Holo_Dis}
\end{document}